\documentclass[twocolumn]{aastex62}
\usepackage{CJK}
\usepackage{amsmath}
\usepackage{url}

\newcommand{\dd}{\mathrm{d}}

\graphicspath{{./}{figures/}}

\received{}
\revised{}
\accepted{}
\submitjournal{ApJ}

\shorttitle{Modeling quasar light curves with an autoencoder}
\shortauthors{Tachibana et al.}

\begin{document}
\begin{CJK*}{UTF8}{gbsn}

\title{Deep modeling of quasar variability}

\correspondingauthor{Matthew J. Graham}
\email{mjg@caltech.edu}

\author[0000-0001-6584-6945]{Yutaro~Tachibana~(優太朗橘)}
\altaffiliation{These authors contributed equally to this work}
\affil{Department of Physics, Tokyo Institute of Technology, 2-12-1 Ookayama, Meguro-ku, Tokyo 152-8551,
Japan}
\affil{Department of Physics, Math, and Astronomy, California Institute of Technology, Pasadena, CA, 91125, USA}

\author[0000-0002-3168-0139]{Matthew~J.~Graham}
\altaffiliation{These authors contributed equally to this work}
\affil{Department of Physics, Math, and Astronomy, California Institute of Technology, Pasadena, CA, 91125, USA}

\author{Nobuyuki~Kawai}
\affil{Department of Physics, Tokyo Institute of Technology, 2-12-1 Ookayama, Meguro-ku, Tokyo 152-8551,
Japan}

\author{S.~G.~Djorgovski}
\affil{Department of Physics, Math, and Astronomy, California Institute of Technology, Pasadena, CA, 91125, USA}

\author{Andrew~J.~Drake}
\affil{Department of Physics, Math, and Astronomy, California Institute of Technology, Pasadena, CA, 91125, USA}

\author{Ashish~A.~Mahabal}
\affil{Department of Physics, Math, and Astronomy, California Institute of Technology, Pasadena, CA, 91125, USA}

\author{Daniel~Stern}
\affil{Jet Propulsion Laboratory, California Institute of Technology, 4800 Oak Grove Drive, Pasadena, CA 91109, USA}



\begin{abstract}
Quasars have long been known as intrinsically variable sources, 
but the physical mechanism underlying the temporal optical/UV variability is still not well understood.  
We propose a novel nonparametric method for modeling and forecasting 
the optical variability of quasars utilizing an autoencoder neural network
to gain insight into the underlying processes. 
The autoencoder is trained with $\sim$15,000 decade-long quasar light curves 
obtained by the Catalina Real-time Transient Survey selected with negligible flux contamination 
from the host galaxy. The autoencoder's performance in forecasting the temporal flux variation of quasars is superior to 
that of the damped random walk process.  We find a temporal asymmetry in the optical variability and
a novel relation -- the amplitude of the variability asymmetry 
decreases as luminosity and/or black hole mass increases -- is suggested with the help of autoencoded features. 
The characteristics of the variability asymmetry are
in agreement with those from the self-organized disk instability model, 
which predicts that the magnitude of the variability asymmetry decreases 
as the ratio of the diffusion mass to inflow mass in the accretion disk increases. 
\end{abstract}



\section{Introduction} \label{sec:intro}
Quasars are a key population for investigating and understanding 
the physics of accretion of matter under extreme physical conditions. 
Several hundred thousand quasars have been spectroscopically confirmed so far and 
many attempts have been made to determine the characteristics of their temporal flux variability. 
However, the physical mechanisms underlying the variability remain poorly understood, in part  
due to the difficulty in parameterizing its aperiodicity. 

In the optical/UV, it is only the variability amplitude and its correlation with timescale that have so far been suggested to be related to intrinsic physical parameters.
For example, the amplitude of quasar optical variability increases with decreasing luminosity, restframe wavelength, 
and Eddington ratio (e.g., \citealt{Wills93,Giveon99,Vanden04}), 
and the structure function tends to possess a steeper slope for quasars with a larger black hole mass \citep{Caplar17}. The correlation with black hole mass is still unclear, however, with different studies advocating either positive or negative relationships
(e.g., \citealt{Wold07, Kelly09, Zuo12}), 
depending on the degree to which observational biases have been eliminated.
Physical mechanisms underlying the optical/UV variability 
have been proposed: 
the superposition of supernovae \citep{Aretxaga97, Kawaguchi98}, 
microlensing \citep{Hawkins93, Hawkins10}, 
thermal fluctuations from magnetic field turbulence \citep{King04, Kelly09, Kelly11}, 
and instabilities in the accretion disk \citep{Takeuchi95, Kawaguchi98}.

Recently a large attempt has been made to reveal the latent physical process underlying extremely large flux variations ($\Delta m$ $\gtrsim$ 1 mag) in quasars. 
Tidal disruption events (TDEs), 
large amplitude microlensing, 
a large change of obscuration or accretion rate, and 
supernovae have been
proposed for such extreme temporal variabilities 
(e.g., \citealt{Meusinger10, Drake11, Bruce16, Lawrence16, Ruan16, Graham17, Stern18, Ross18, Assef18}), 
but it remains unclear whether or how they relate to the more general optical variability seen in quasars.

To describe quasar optical variability,  \cite{Kelly09} proposed a continuous time first-order autoregressive model, also known as the
Ornstein-Uhlenbeck or damped random walk (DRW) process,  which is a particular type of Gaussian process characterized by two parameters: 
$\tau$, the relaxation time,  and $\sigma$, the variability on timescales much shorter than $\tau$. Several authors have shown that the DRW
process provides a better statistical model for most quasar variability when compared to a range of alternative stochastic/deterministic models (e.g., \citealt{Andrae13}). However, \cite{Kozlowski17}, pointed out that the best-fit DRW processes are biased in $\tau$
due to an insufficient temporal baseline in existing surveys for probing the white noise portion of the power spectral density (PSD). This paper shows 
that a temporal baseline at least ten times longer than $\tau$ is necessary to properly constrain $\tau$. Any reported correlations between these model parameters and physical parameters, such as black hole mass or Eddington ratio, are therefore potentially analysis artifacts. 
Additionally, deviations from a DRW process in quasar variability have begun to be recognized. 
\textit{Kepler} light curves with $\sim$30 min sampling revealed a steeper power-law index
of about $-3$ at very high frequency (less than a few months; e.g., \citealt{Mushotzky11,Kasliwal15}), 
which is a significant deviation from the DRW process. 
On very long timescales (at lower frequencies than the typical timescale of a quasar light curve), 
\cite{Guo17} found that the observed residual scatter in $\sigma$ 
is too large for uncertainties in the DRW process parameter derived from 1,678 light curves of low redshift quasars with low black hole mass. 
They also suggested that the scatter can be explained if the low frequency PSD slope is about $-1.3$. 
\cite{Mushotzky11} concluded that individual quasars exhibit intrinsically different PSD slopes, 
indicating that the DRW process is too simplistic to describe optical quasar variability (e.g., \citealt{Graham14, Kasliwal15, Caplar17}).
The situation would likely be even worse for more complex stochastic models. More phenomenological parameters 
would be even more difficult to connect with underlying physical processes.

In this work, we present an initial application of the autoencoder, which is a type of unsupervised (deep) machine learning algorithm,
 to quasar temporal flux behavior 
by assuming that quasar temporal variability can be represented in a low dimensional space.  
The training and the validation of the model is performed with quasar light curves obtained by the Catalina Real-time Transient Survey 
(CRTS;\footnote{\url{http://crts.caltech.edu}} \citealt{Drake09, Mahabal11}), 
which is the largest open (publicly accessible) time domain survey currently available.
The representative expressions or characterizing features of temporal variability are acquired by the autoencoder itself in an unsupervised way, 
and thus modeling and forecasting is performed without any prior assumptions. We also propose a methodology for associating the representative expressions (autoencoded features; AE features) 
with physical parameters utilizing a simple multilayer perceptron (MLP) and then show its validity. 

This paper is structured as follows: in section 2, we describe the method and data selection and in section 3, the results of applying the autoencoder to extract features and to forecast quasar variability. Section 4 discusses the features and their relation to physical parameters and models.  Section 5 presents our conclusions. Alongside this paper, the scripts used for the analysis shown in this work are available online\footnote{ 
\url{https://github.com/yutarotachibana/CatalinaQSO_AutoEncoder}}. 

\section{Method}
In this section, we discuss CRTS, the photometric calibration method employed by the pipeline of the survey project, 
the data selection criteria we employ in this work, and the basic structure of the autoencoder we use to model and forecast quasar variability. 

\subsection{Catalina Real-time Transient Survey}\label{sec:cat}
The CRTS archive\footnote{http://catalinadata.org} contains the Catalina Sky Survey data streams from three telescopes --  the 0.7 m Catalina Sky Survey (CSS) Schmidt and 1.5 m Mount Lemmon Survey (MLS) telescopes in Arizona, and  the 0.5 m Siding Springs Survey (SSS) Schmidt in Australia. These surveys, operated by the Lunar and Planetary Laboratory at the University of Arizona, were designed to search for near-Earth objects, but have proven extremely valuable for astrophysics topics ranging from Galactic transients \citep{drake14} to distant quasars \citep{Graham14,Graham15b,Graham17}. CRTS covers up to $\sim$2500 deg$^2$ per night, with 4 exposures per visit, separated by 10 min. The survey observes over 21 nights per lunation. The data are broadly calibrated to Johnson $V$ (see \citealt{drake13} for details) and the current CRTS data set contains time series for approximately 400 million sources to $V \sim 20$ above Dec $> -30$ from 2003 to 2016 May (observed with CSS and MLS) and 100 million sources to $V \sim 19$ in the southern sky ($-75 < $Dec$ < 0$) from 2005 to 2013 (from SSS).

There are few data sets with sufficient sky coverage, temporal coverage, and sampling to enable us to investigate quasar optical variability systematically. 
The largest data sets which can be used for research on the long term optical variability of quasars currently
are SDSS with POSS, Pan-STARRS1 (e.g., \citealt{MacLeod12, Morganson14}), 
and CRTS. Among these, CRTS provides the best dataset for investigating the temporal flux variation
on timescales from weeks to decades 
due to its large number of objects and observation cadence. 

\begin{figure}[t] 
 \begin{center}
  \includegraphics[width=\linewidth]{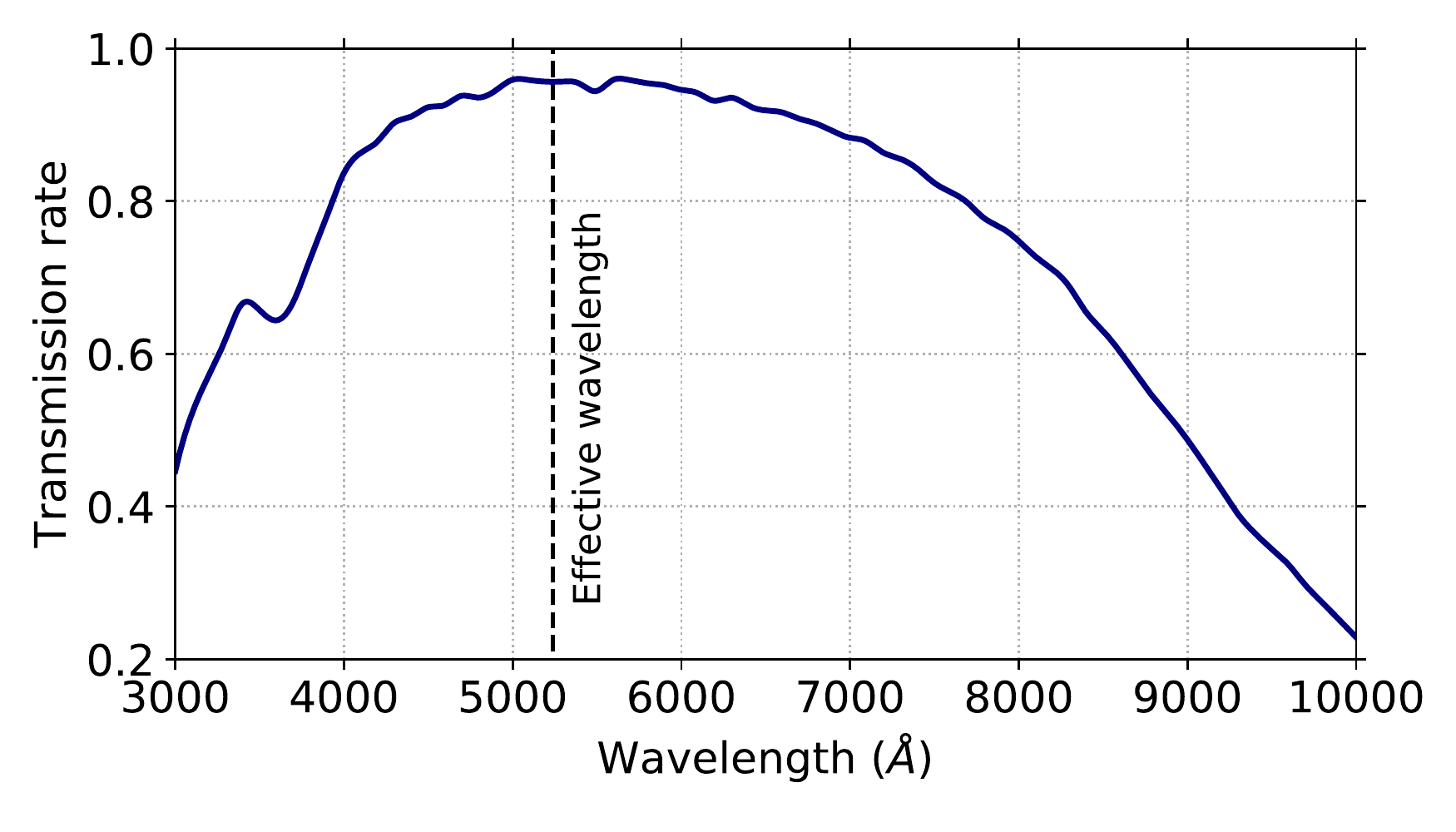}
 \end{center}
 \caption{The transmission curve of the unfiltered system employed by CRTS. 
 The effective wavelength is represented by the vertical dashed line. }  
 \label{fig:tr}
\end{figure}

The error model used for CRTS is incorrect: errors at the brighter magnitudes are overestimated and those at fainter magnitudes ($V > 18$) are underestimated \citep{palaversa13, drake14}. In this analysis, we employ the improved error model derived in \cite{2017MNRAS.470.4112G}; the actual CRTS error model will be fixed in a future release. We apply the same preprocessing steps described in \cite{Graham15b} to all light curves, which remove outlier photometric points and combine all exposures for a given night to give a single weighted value for that night. We also remove sources associated with nearby bright stars or identifiable as blends from a combined multimodality in their magnitude and observation position, i.e., the spatial distribution of all points in a light curve is best described by $n > 1$ Gaussians.  

\subsection{Data Selection}
\begin{figure*}[t] 
 \begin{center}
  \includegraphics[width=\linewidth]{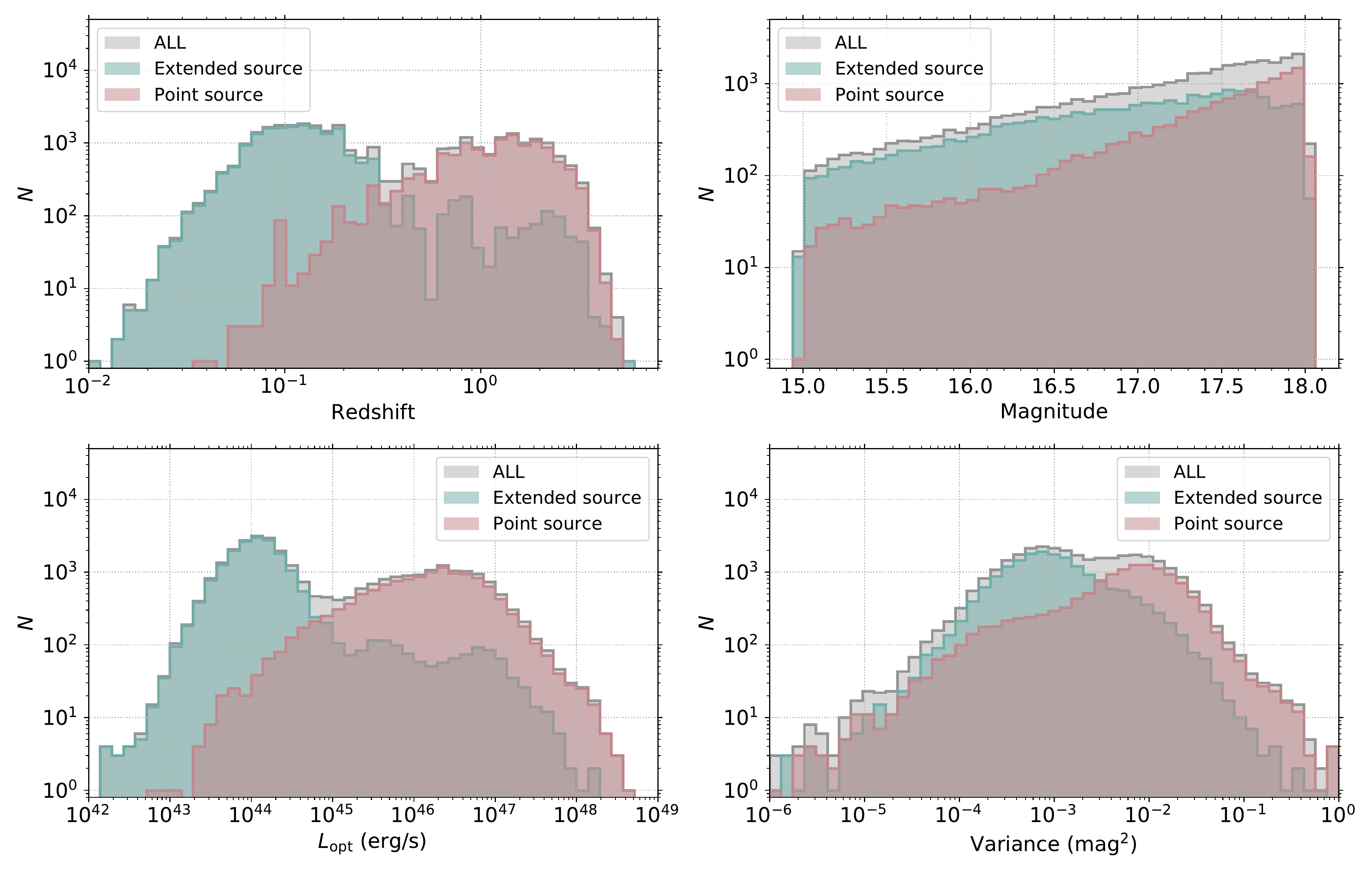}
 \end{center}
 \caption{The distribution of redshift (upper left), average magnitude (upper right), 
 optical luminosity (lower left), and intrinsic variance (lower right). 
 Green and red indicate extended sources (resolved on the PS1 image) 
 and point sources (unresolved on the PS1 image), respectively.
 The sum of them are denoted by the grey histogram in each panel. }  
 \label{fig:hist_cat}
\end{figure*}

We have crossmatched 555,692 sources classed as ``QSO" in SDSS DR15 \citep{Aguado2019} against the CRTS data set with a 3'' matching radius. We selected objects within the magnitude range $15 \le V \le 18$ to minimize systematic effects from error estimation and saturation and excluded known blazars. 40,736 spectroscopically confirmed quasars lie within these ranges in CRTS.

To check the characteristics of the quasars, 
we calculated the variance of the light curves and the optical luminosities, 
where the intrinsic variance is referred to as the variance in this paper, and is described by:
\begin{align}
\sigma_{\mathrm{mag}}^2 = \frac{1}{N-1}\sum_{i=1}^{N}(mag_i - \overline{mag})^2 - 
	\frac{1}{N}\sum_{i=1}^{N}e_i^2. 
\end{align}
where $N$ is the number of data points, $mag$ and $e$ are the observed magnitude and its uncertainty, respectively, 
and $\overline{mag}$ is the weighted average of the magnitudes. 
The optical luminosity ($L_{\mathrm{opt}}$) can be approximately calculated by: 
\begin{align}
L_{\mathrm{opt}} = 
	4\pi D_{\mathrm{L}}^2 F_0\lambda_{\mathrm{eff}} 
    \times 10^{-(\overline{mag} -A_{\mathrm{crts}})/2.5} ~ \mathrm{erg s^{-1}}, 
\end{align}
where $D_{\mathrm{L}}$ is the luminosity distance calculated with 
$\Omega_{\Lambda} = 0.728$, $\Omega_{\mathrm{M}} = 0.272$, 
and $H_0 = 70.4\ \mathrm{km}\ \mathrm{s}^{-1}\mathrm{Mpc}^{-1}$ \citep{Jarosik11}, 
$F_0 = 3.968\times10^{-9} \ \mathrm{erg}\ \mathrm{cm}^{-2}\mathrm{s}^{-1}$\AA$^{-1}$  
is the zero point flux density,\footnote{\url{http://svo2.cab.inta-csic.es/svo/theory/fps3/index.php?id=Misc/CRTS.C}} 
$\lambda_{\mathrm{eff}} = 5237.44$ \AA\  is the effective wavelength of the CRTS filter system (see Fig.~\ref{fig:tr}),  
and $A_{\mathrm{crts}}$ is the Galactic absorption at the effective wavelength along the line-of-sight. 
The CRTS Galactic absorption is estimated based on 
the total extinction in the $V$-band provided by IRSA,\footnote{\url{https://irsa.ipac.caltech.edu/applications/DUST/}} 
obtained by using the Python package 
\texttt{astroquery}.\footnote{\url{https://astroquery.readthedocs.io/en/latest/irsa/irsa_dust.html}}
The extinction in the $V$-band can be translated to that at  $\lambda_{\mathrm{eff}}$ 
through an empirical relation between \AA\ and $A_{\lambda}/A_{\mathrm{{V}}}$ 
given by \cite{O'Donnell94}, where we adopt $R_{\mathrm{V}}=3.1$. 
The \texttt{extinction} package\footnote{\url{https://extinction.readthedocs.io/en/latest/\#extinction}} 
is used for converting $A_{\mathrm{V}}$ to $A_{\mathrm{crts}}$.
Fig.~\ref{fig:hist_cat} shows the distributions of redshift, mean magnitude, 
optical luminosity, and variance for the quasar sample.
The histograms colored by green and red overplotted on the gray histograms 
indicate the distribution of sources identified as an extended source or as a point source in their respective PS1 image
\citep{Tachibana18}.\footnote{Strictly speaking, a PS1 counterpart within 1 arcsec from a CRTS quasar. } 
One can see that there are two obvious classes in the data set: 
(1) resolved, nearby, intrinsically fainter, and lower variable sources, and 
(2) unresolved, far away, intrinsically brighter, and higher variable sources. 
\begin{figure}[t] 
 \begin{center}
  \includegraphics[width=\linewidth]{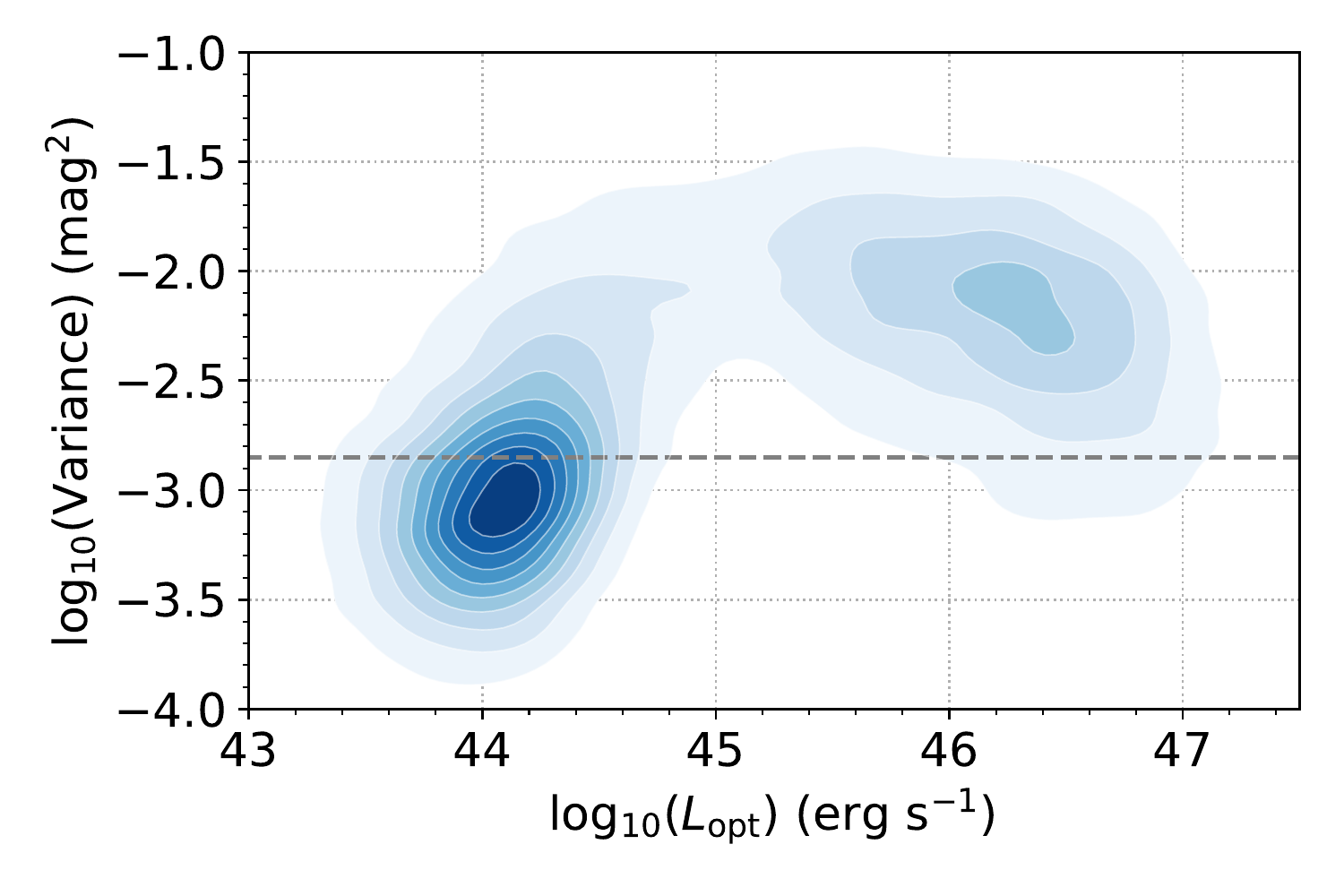}
 \end{center}
 \caption{The distribution of our original CRTS quasar catalog sample 
 on the $L_{\mathrm{opt}}$-$\sigma_{\mathrm{mag}}^2$ plane. 
 The dashed line indicates the typical uncertainty in flux measurements. Contour lines indicate
 the 10th - 90th percentiles of the distribution.}  
 \label{fig:temporal_feat}
\end{figure}
The two groups are clearly distinguishable 
in the $L_{\mathrm{opt}}$--$\sigma^2_{\mathrm{mag}}$ plane, 
as shown in Fig.~\ref{fig:temporal_feat}. 
This behavior can be interpreted as the combination of flux coming from 
the stable host galaxy and the partially-visible variable accretion disk.
Significant flux contamination at the faint end is unavoidable as 
CRTS measurements are produced by aperture photometry (see Section \ref{sec:cat}) and so 
a part of the resolved host galaxy must be inside the aperture used. 
For sources brighter than $L_{\mathrm{opt}} \sim 10^{45} \ \mathrm{erg s^{-1}}$, 
the variance and luminosity are anticorrelated, 
which is consistent with previous research mentioned in Section \ref{sec:intro}. 
We are therefore able to identify sources showing variability purely originating from the disk
with $L_{\mathrm{opt}}>10^{45} \ \mathrm{erg s^{-1}}$.

15,438 quasars were selected, which should contain minimal 
flux contamination from the host galaxy.\footnote{In addition to the luminosity threshold, 
sources with $10^{-4}\ \mathrm{mag}^2 < \sigma_{\mathrm{mag}}^2 < 10^{-1}\ \mathrm{mag}^2$, 
$T_{\mathrm{obs}} > 2500$ d, and
$n_{\mathrm{obs}} > 50$, 
are selected, where $T_{\mathrm{obs}}$ and  $n_{\mathrm{obs}}$ 
is the observation length and the number of observations, respectively.} 
This selection is crucial to investigate quasar variability, namely disk variability,
because the contamination significantly suppresses the variation amplitude at its faint state 
and we cannot subtract the contamination from the total brightness 
as we do not know the true flux level of the host galaxy. 

\subsection{Simulated Light Curves}
\begin{figure*}[t!] 
 \begin{center}
  \includegraphics[width=\linewidth]{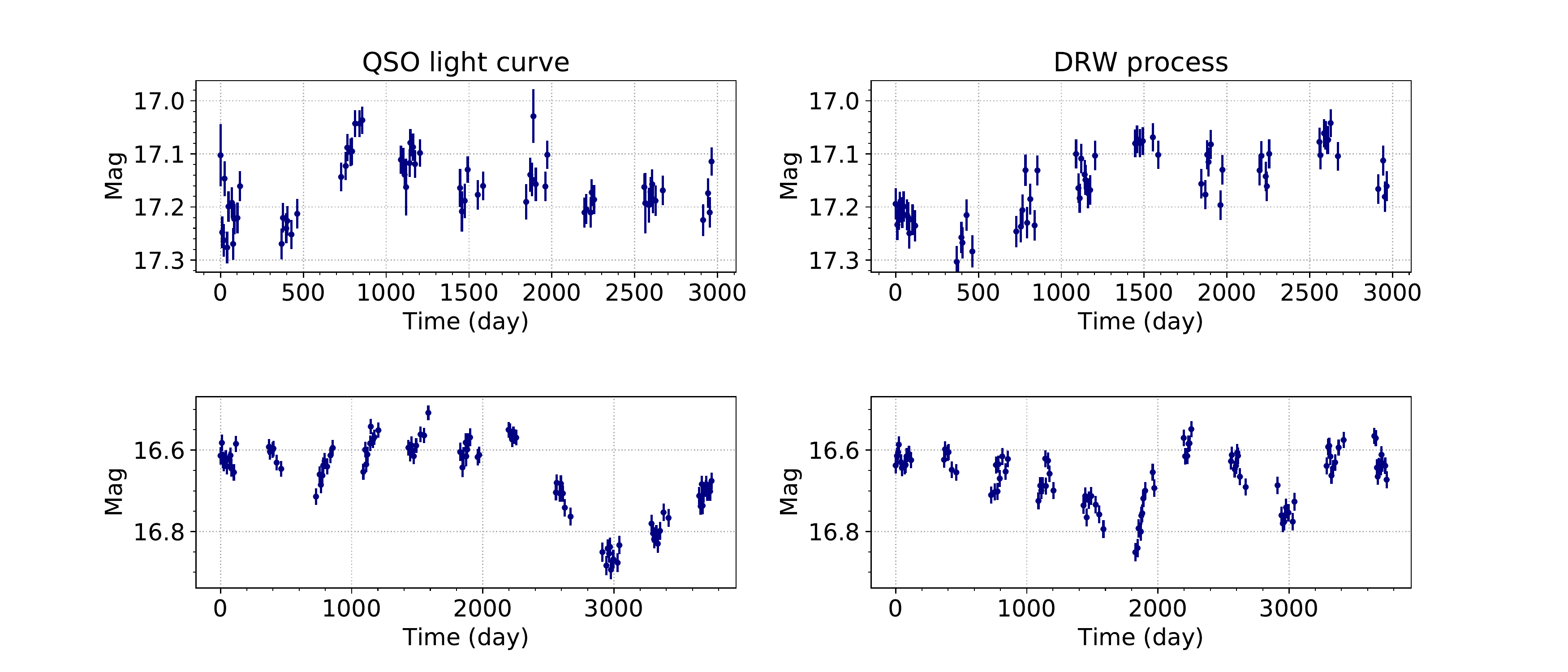}
 \end{center}
 \caption{Examples of quasar light curves (left) and simulated light curves (right) 
 generated by the DRW process with the same observation cadence 
 and same parameters $b$, $\sigma$, and $\tau$ 
 as the associated quasar light curve. 
 The modeled error is added to the DRW process as the measurement uncertainties. }  
 \label{fig:mock_lcs}
\end{figure*}
Simulated light curves are commonly used to assess systematic biases
because observational biases such as observation gaps (i.e., the window function of the observation), 
which can generate systematic and puzzling results (e.g., \citealt{Suberlak17}), 
should show in analysis results for both the real and the simulated data. 
Since the expected behavior for optical quasar variability is 
that it approximately follows a DRW process (see Section \ref{sec:intro}), 
we generate simulated light curves using the actual observation times, but replacing 
the observed magnitudes with expected values under a DRW process. 

Formally, the temporal behavior of a DRW process $X(t)$ is given by:
\begin{align} \label{eq:ou}
\dd X(t) = - \frac{1}{\tau} X(t) \dd t + \sigma \sqrt{\dd t} \epsilon(t) + b \dd t, 
\end{align}
where $\epsilon(t)$ is a white noise process with zero mean and variance equal to 1 
and $b = \overline{X(t)}/\tau$. 
The corresponding likelihood function involves an exponential covariance matrix: 
\begin{align}
S_{ij} = \frac{\tau\sigma^2}{2}\exp(-|t_i - t_j|/\tau). 
\end{align}
The model parameters for the simulated light curves, $b$, $\sigma$ and $\tau$, 
are the same as those derived from the DRW process fit to the associated quasar light curve.
In addition, we added a Gaussian deviate derived from the 
empirical function:
\begin{align}
e_{\mathrm{mag}} = a\exp(b\times mag)+c
\end{align}
fit to the quasar dataset, and the modeled error is treated as the measurement uncertainty on the simulated light curves. 
Note that both fitting and simulation is in the quasar restframe.

Examples of observed and simulated quasar light curves are displayed in Fig.~\ref{fig:mock_lcs}. 
The interpretation of analysis results is performed by comparison between the results for the two data sets.

\subsection{Autoencoder Neural Network}\label{sec:aea}
\begin{figure}[t] 
 \begin{center}
  \includegraphics[width=\linewidth]{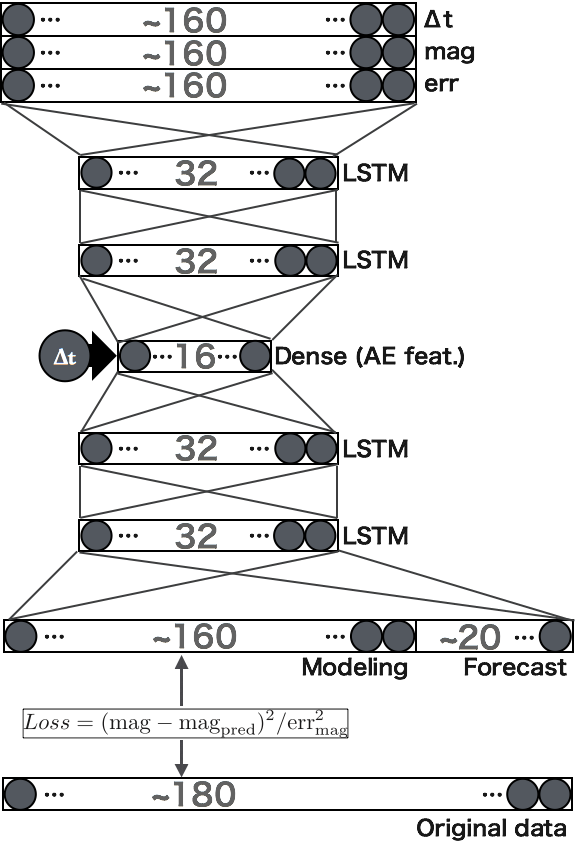}
 \end{center}
 \caption{Diagram of the RNN autoencoder architecture constructed for modeling quasar light curves 
 	in this work.  See \S\ref{sec:aea} for details.}  
 \label{fig:aea}
\end{figure}
An autoencoder is a type of unsupervised neural network which is trained to reconstruct the original input while compressing the data in the process so as to discover a more efficient and reduced representation in an internal (hidden) layer. The main purpose of this architecture is dimension reduction and as the number of nodes in the hidden layer is smaller than in the input and output layers, fundamental information should be 
condensed at the layer with the smallest number of nodes. This architecture facilitates classification and also optimum modeling of the input data. 

For sequence-to-sequence data, the autoencoder can be implemented using a recurrent neural network \citep[RNN; see][for a review]{Lipton2015} architecture. Traditional neural networks assume that all inputs (and outputs) are independent of each other but RNNs perform the same task for every element of a sequence with the output at a particular timestep forming part of the input to the next timestep. This means that information is retained about what has been calculated so far and this can affect the current calculation and prediction. RNNs have been used in astronomy for time series classification \citep{Charnock17,Naul18,Becker20}. The RNN autoencoder network is trained with time series as input to reproduce the same time series as the output. 
The coded representation in the hidden layer is thus a time-dependent compression and can be interpreted as features of the input time series. 
With these features, \cite{Naul18} demonstrated that the accuracy of supervised variable star classification 
is superior to or at least consistent with that of a classifier with expert-chosen hand-selected features.

The autoencoder neural network that we constructed for modeling and forecasting quasar light curves
is displayed in Fig.~\ref{fig:aea}. This network uses two LSTM\footnote{Long short time memory (LSTM) 
is a type of RNN;  for detailed information about LSTMs, see \cite{Jain1999}.} 
layers of size 32 for encoding (reducing the input) and two for decoding (reconstructing the input), with an 
autoencoded feature size of 16 (AE features hereafter). We input the measurement values, the differences between sampling times $\Delta t$ (to deal with the irregular time sampling of the data), and the measurement errors. Since we are also interested in forecasting, we have excluded the last 500 days of data for each source. The AE features are constructed by passing the output of the 
last recurrent encoding layer into a single fully-connected layer with a linear 
activation function and the desired output size. 
The decoder repeats the AE features $N_{\mathrm{T}}$ times, 
where $N_{\mathrm{T}}$ is the length of the next layer, 32 in this architecture, 
and then appends the $\Delta t$ values to the corresponding elements of the resulting vector sequence. 
The decoder network is constructed from another series of LSTM layers, 
with a final linear layer to generate the original light curve, 
i.e., the output is 500 days longer than the input data.  
The model, therefore, performs modeling and forecasting simultaneously. 
The loss (weighted mean squared error) is defined by:
\begin{align}\label{eq:redchisq}
\mathrm{loss} = \frac{1}{N_{\mathrm{T}}} \sum_{i=1}^{N} \sum_{j=1}^{N_\mathrm{T}} 
	\left( \frac{mag_i^{(j)} - \widehat{mag}_i^{(j)}} {\sigma_i^{(j)}} \right)^2, 
\end{align}
where $N$ is the number of light curves, and $mag_i^{(j)}$, $\widehat{mag}_i^{(j)}$, 
and $\sigma_i^{(j)}$ are the $j$th measurement, reconstruction value, and measurement error 
of the $i$th light curve, respectively; this
reduces the penalty for reconstruction errors when the measurement error is large. 
We also apply a 25\% dropout between LSTM layers to generalize the ability 
to model and forecast quasar light curves. We note that the architectural hyperparameters of the network, i.e., 
the number of layers, number of nodes per layer, number of nodes in the hidden layer, etc., are arbitrarily chosen 
to provide a network similar to the one employed by \cite{Naul18}. Bayesian optimization of these quantities is 
possible but can be computationally expensive and by experimentation we found that the results of the network 
were robust to changes by factors of two in the values used here.

\section{Result}
\subsection{Training the Autoencoder}
\begin{figure*}[t] 
 \begin{center}
  \includegraphics[width=150mm]{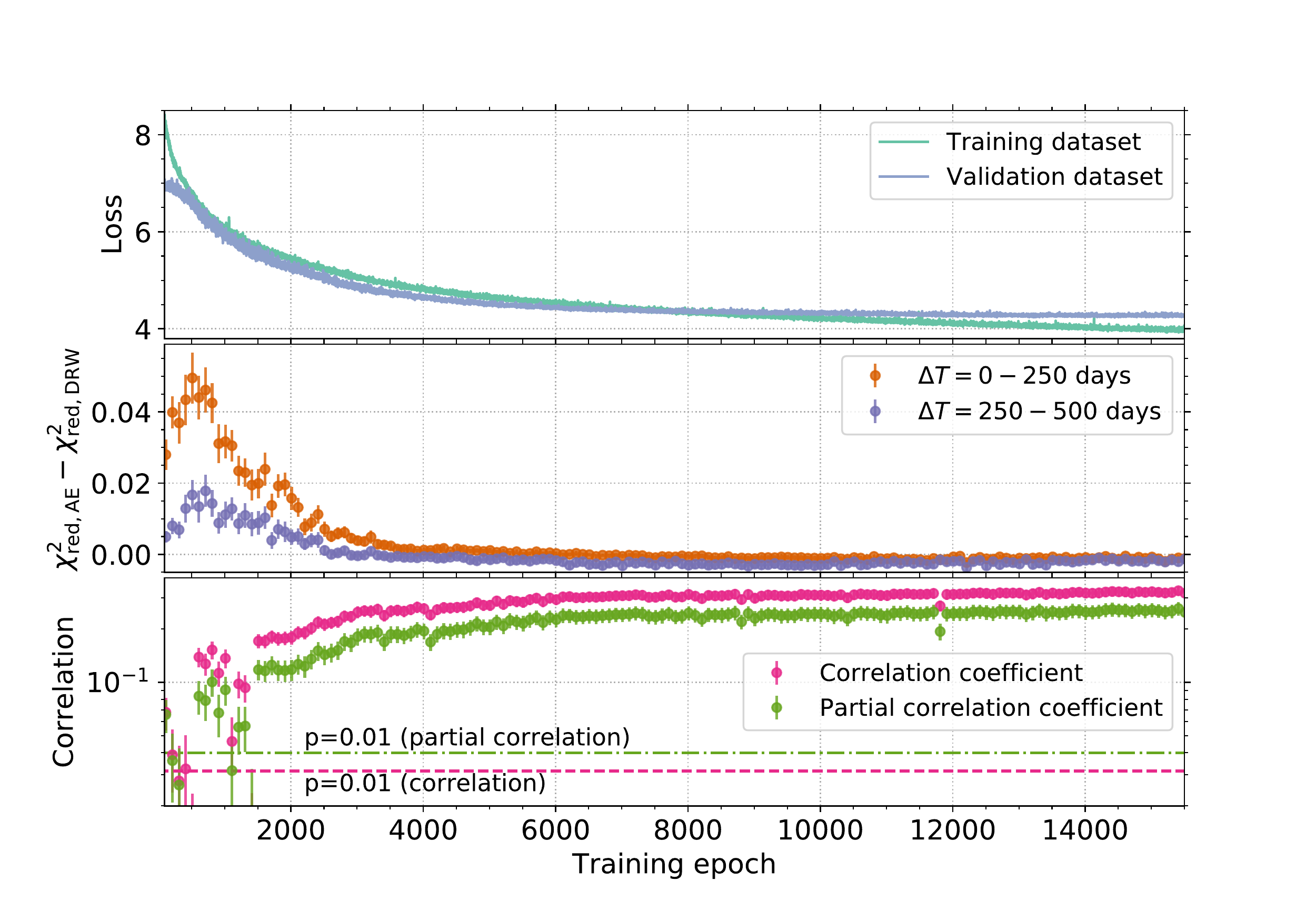}
 \end{center}
 \caption{The reconstruction loss for the training data set and for the validation data set (top), 
 the $\chi_{\mathrm{red}}^2$ value for $\Delta T_{\mathrm{pred}}$ = 0--250 d to 250--500 d 
 (middle; for the definition of $\Delta T$, see text), 
 and the (partial) correlation coefficient with $L_{\mathrm{opt}}$
 as a function of training epoch (bottom). 
 The correlation coefficient and the partial correlation coefficient corresponding to $p$-value $=0.01$ is 
 represented by the pink dashed line and the green dot-dashed line, respectively.  }  
 \label{fig:tr_seq}
\end{figure*}
To train and validate the autoencoder shown in Fig.~\ref{fig:aea}, 
we divided the quasar dataset into a training dataset (80\%; 12,350 sources) 
and a validation dataset (20\%; 3,088 sources). 
The input ($mag_{\mathrm{in}}$) and the target magnitudes ($mag_{\mathrm{tar}}$) 
are normalized by the average $\overline{mag_{\mathrm{in}}}$ 
and the standard deviation $\sigma_{mag_{\mathrm{in}}}$ of the input magnitude; 
\begin{align} \label{eq:norm}
\hat{mag}_{\mathrm{in}} &= (mag_{\mathrm{in}} - \overline{mag_{\mathrm{in}}})/\sigma_{mag_{\mathrm{in}}} \\ 
\hat{mag}_{\mathrm{tar}} &= (mag_{\mathrm{tar}} - \overline{mag_{\mathrm{in}}})/\sigma_{mag_{\mathrm{in}}}. 
\end{align}
Also for $\Delta T_{\mathrm{in, tar}}$ and $err_{\mathrm{in, tar}}$, the normalizations 
$\hat{\Delta T}_{\mathrm{in, tar}} = \Delta T_{\mathrm{in, tar}}/365$ and 
$\hat{err}_{\mathrm{in, tar}} = err_{\mathrm{in, tar}}/\sigma_{x_{\mathrm{in}}}$ 
are applied. 
We note that the inputs do not have any information 
on the forecasting part (the last 500 days) as we used only  
$\overline{mag_{\mathrm{in}}}$ and $\sigma_{mag_{\mathrm{in}}}$
for the normalization of both the input and the output. 

Fig.~\ref{fig:tr_seq} shows the loss (see  eqn.~\ref{eq:redchisq}) for the training dataset and the validation data set. 
We used \texttt{Adam} optimization \citep{Kingma14} 
with standard parameter values $\beta_1 = 0.9$, $\beta_2 = 0.999$,
a learning rate of $\eta = 1\times10^{-4}$ and a batch size of 256. 
All models are implemented with the \texttt{Keras} 
package.\footnote{\url{https://keras.io/}} 
The top panel in Fig.~\ref{fig:tr_seq} shows that both the validation loss and the training loss decrease
as a function of the training epoch. 
While the training loss and the validation loss values cross at the training epoch of $\sim$8000, 
no obvious signal of overfitting is seen. 
The final loss for the validation dataset is $\sim$4.25, 
which might seem somewhat large for a reduced chi-square $\chi_{\mathrm{red}}^2$, 
but is acceptable as the loss is calculated for both the modeling part and the forecasting part of the output. 

The middle panel in Fig.~\ref{fig:tr_seq} shows the forecasting accuracy 
evaluated from the difference of the reduced chi-square ($\chi_{\mathrm{red}}^2$; see Eq.(\ref{eq:redchisq})) 
of the 
autoencoder model (AE model; $\chi_{\mathrm{red, AE}}^2$) 
and the DRW process model (DRW model; $\chi_{\mathrm{red, DRW}}^2$, see Eq.(\ref{eq:ou_pred})) 
for time ranges $0\ \mathrm{days} \leq \Delta T_{\mathrm{pred}} < 250 \ \mathrm{days}$ 
and $250\ \mathrm{days} \leq \Delta T_{\mathrm{pred}} < 500 \ \mathrm{days}$, 
where $\Delta T_{\mathrm{pred}}$ is 
the time difference from the beginning of the forecasting part of the output. 
$0\ \mathrm{d} \leq \Delta T_{\mathrm{pred}} < 250 \ \mathrm{d}$ and 
$250\ \mathrm{d} \leq \Delta T_{\mathrm{pred}} < 500 \ \mathrm{d}$
is thus the first half and the latter half of the forecasting part, respectively. 
Since the accuracy of the AE model is defined by $\chi_{\mathrm{red, AE}}^2 - \chi_{\mathrm{red, DRW}}^2$, 
a smaller value indicates a higher accuracy. 
As shown in Fig.~\ref{fig:tr_seq}, the forecasting accuracy increases as the training proceeds. 

In addition, we have confirmed that the AE features actually acquire information on physical parameters as training proceeds. 
The bottom panel in Fig.~\ref{fig:tr_seq} shows 
the correlation coefficient and the partial correlation coefficient 
between the AE features and optical luminosity, 
where these values are calculated on the validation data set. 
For the partial correlation coefficient, the variance of the light curve, 
which is known to be correlated with optical luminosity,  
is considered to be a latent variable, and its effect is removed from the 
correlation coefficient 
(see Section \ref{sec:aefeat} for the method to calculate the correlation coefficient 
between the AE features and a physical parameter). 
The partial correlation coefficient can be calculated by:
\begin{align}
\rho_{AE, L_{\mathrm{opt}}\cdot Var} = 
	\frac{ \rho_{AE, L_{\mathrm{opt}}} - \rho_{AE, Var}\rho_{L_{\mathrm{opt}}, Var}}
	{\sqrt{ 1- \rho_{AE, Var}^2 } \sqrt{ 1- \rho_{L_{\mathrm{opt}}, Var }^2}}, 
\end{align}
where $\rho_{x, y}$ is the correlation coefficient between $x$ and $y$, 
and $AE$, $L_{\mathrm{opt}}$, and $Var$ refer to the AE features, the optical luminosity, and the variance of the light curve respectively. 

Both the partial correlation coefficient and the correlation coefficient are not statistically significant (below $p=0.01$) at the beginning of training, 
while after $\sim$500 training epochs, both quantities are significantly above the significance levels. This result is expected but shows the expediency of the autoencoder in modeling the quasar light curves.  Previous work has shown that there is information on quasar physical parameters in their flux variability but extracting it can be involved, e.g., the quantity of interest is the amplitude of variability at a certain time lag in the structure function or the index of the power law fit to it. This result demonstrates that the autoencoder we constructed can automatically acquire such information.

\subsection{Forecasting the Temporal Variability} \label{sec:forecast}
\begin{figure*}[t!] 
 \begin{center}
  \includegraphics[width=\linewidth]{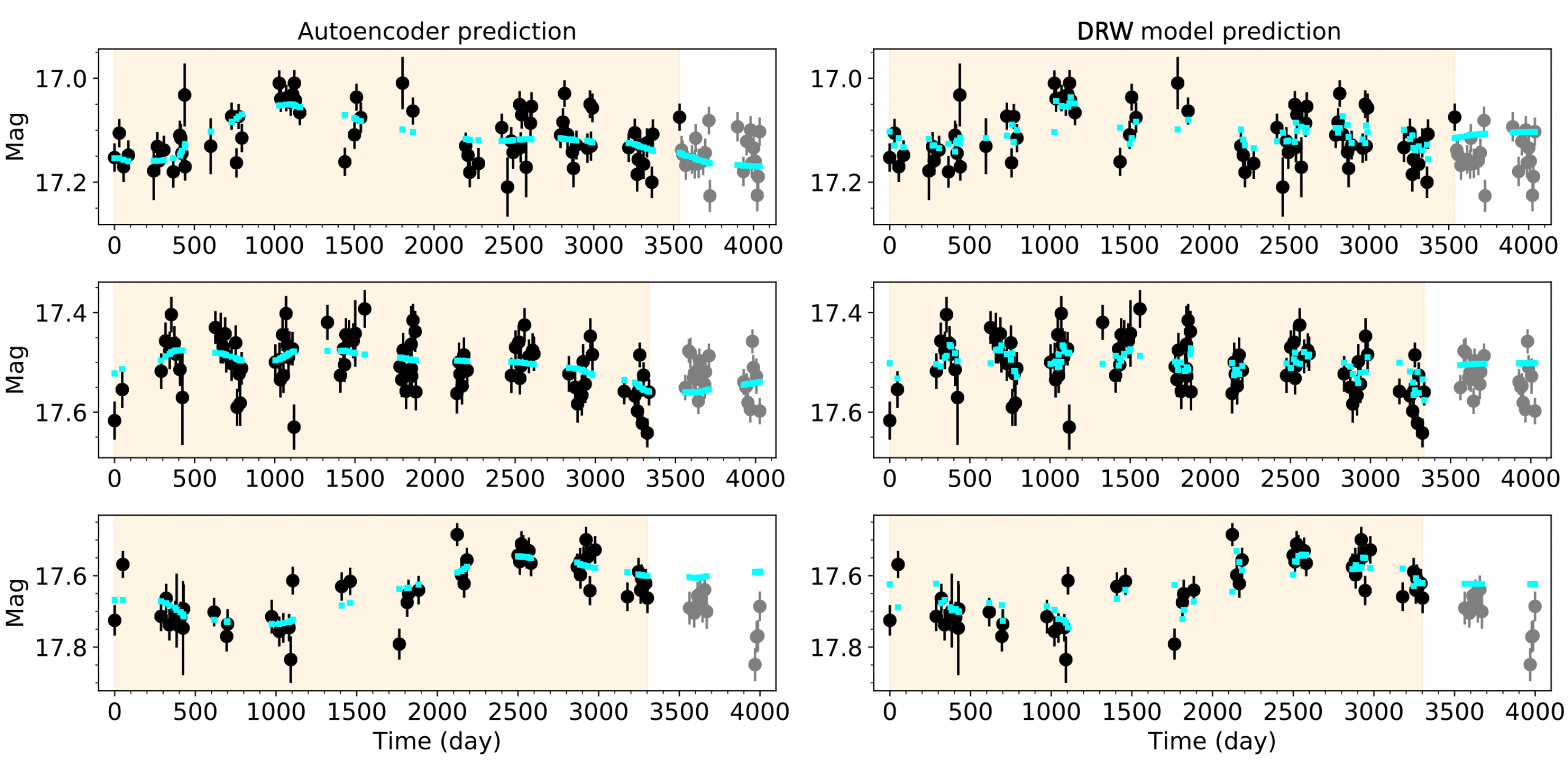}
 \end{center}
 \caption{Examples of modeling and forecasting the quasar light curves by the autoencoder (left) 
 and the DRW process (right). 
 The yellow shadowed region indicates the range fed to the autoencoder, 
 while gray points are the prediction part (last 500 days) which is not used for the input. 
 In the left panels, the cyan squares are the output of the autoencoder. 
 The parameters in the DRW process are calculated by the fit to the input data, 
 where the fitted DRW process is denoted by cyan points in the yellow shadowed region in the right panels, 
 and the expected mean values  with the derived parameters from the  last point of the modeling part 
 are also shown subsequent to the fitted curve.}  
 \label{fig:exam_output}
\end{figure*}
Fig.~\ref{fig:exam_output} shows examples of the output of the autoencoder.
We compare the modeling part and the forecasting part of the autoencoder and the DRW process for the same objects in the left three panels 
and the right three panels, respectively.  The most apparent difference between them is the scatter in short-timescale variability in the modeling part:  the output of the autoencoder is relatively smoother. Short timescale scatter is not resolvable in our data due to the 
sampling cadence and statistical errors. The DRW process, however, includes short-time variability ($\sigma$) to express 
the overall variance of the light curve ($=\tau\sigma^2/2$). In other words, the power law index of the PSD of the DRW
process must be $-2$ above the typical frequency, even if the Fourier power is dominated by noise. The autoencoder, on 
the other hand, models the quasar temporal behavior purely based on the characteristics of the data without any prior 
assumptions. The suppression of such short-time variability in the autoencoder's modeling corresponds to 
a steeper spectral index of the PSD than that of the DRW process in the high frequency regime.

For the forecasting part, the autoencoder seems to output real variations, 
i.e., the output does not fall to the mean value or diverge upward or downward immediately. 
The autoencoder also predicts different behavior to the DRW process. 
We define the prediction of the DRW process as the expectation value from the last data in the modeling part:
\begin{align}\label{eq:ou_pred}
mag(\Delta T_{\mathrm{pred}}) = \mathrm{e}&^{\Delta T_{\mathrm{pred}} / \tau_{\mathrm{in}}} mag(t_{\mathrm{in, last}}) \notag\\ 
	&+ b_{\mathrm{in}}\tau_{\mathrm{in}} (1 - \mathrm{e}^{-\Delta T_{\mathrm{pred}}/\tau_{\mathrm{in}}}), 
\end{align}
where $\tau_{\mathrm{in}}$ and $b_{\mathrm{in}}$ are the DRW process parameters 
derived from fitting the process to the modeling part of the light curve, 
and the $t_{\mathrm{in, last}}$ is the last observation time in the modeling part. 

\begin{figure}[t] 
 \begin{center}
  \includegraphics[width=\linewidth]{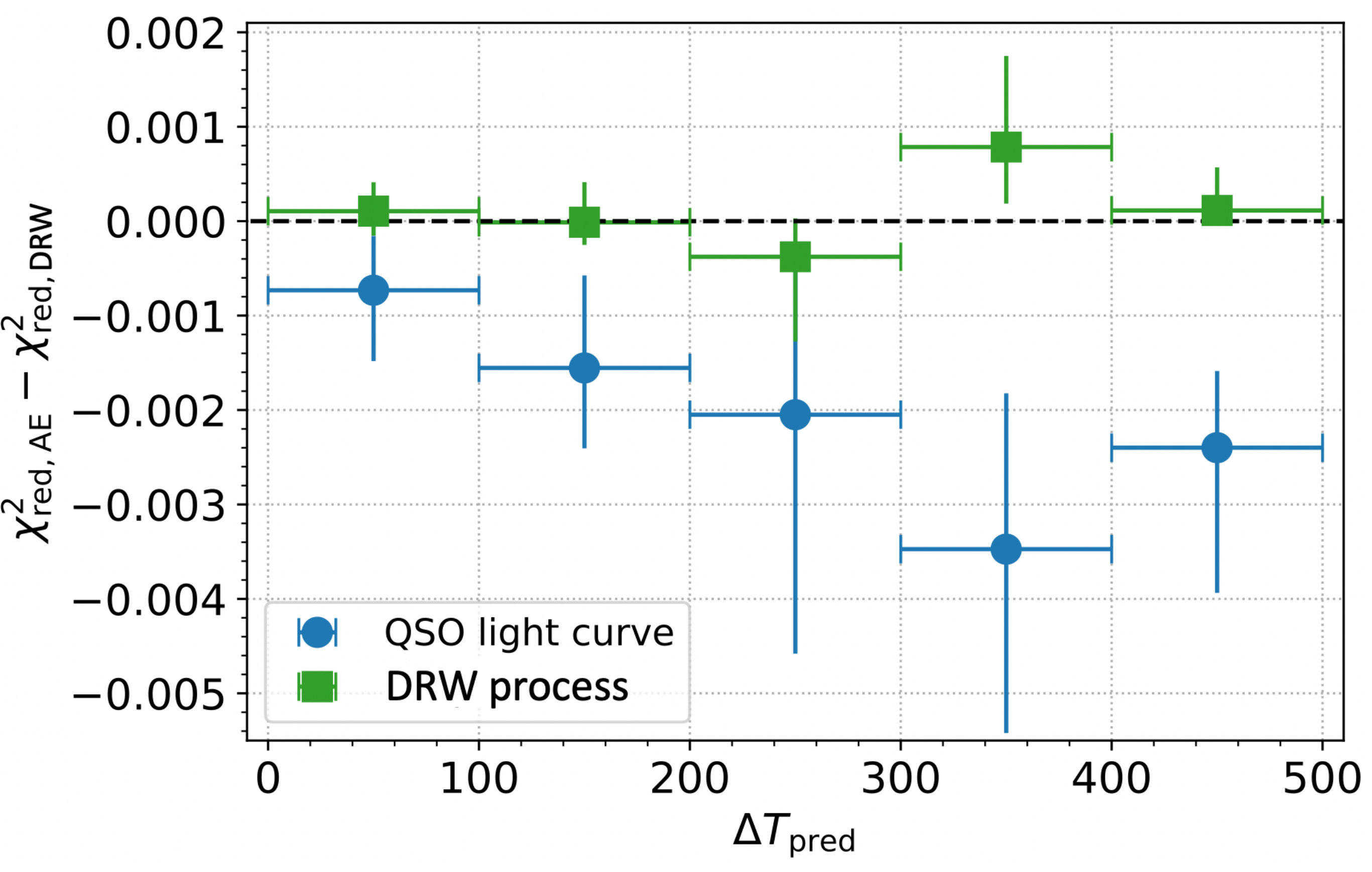}
 \end{center}
 \caption{The difference between the reduced chi square of the AE model $\chi_{\mathrm{red,AE}}^2$ 
 and the DRW model $\chi_{\mathrm{red,DRW}}^2$ as a function of $\Delta T_{\mathrm{pred}}$ 
 for the quasar light curves (blue points) and for the mock light curves (green points). }  
 \label{fig:pred_acc_deltaT}
\end{figure}

To assess the forecasting accuracy of the autoencoder model (AE model), 
we calculated the difference between the reduced chi-square 
for the AE model $\chi_{\mathrm{red, AE}}^2$ and 
the DRW model $\chi_{\mathrm{red, DRW}}^2$ for quasar light curves. 
The blue points in Fig.~\ref{fig:pred_acc_deltaT} show
$\chi_{\mathrm{red, AE}}^2 - \chi_{\mathrm{red, DRW}}^2$
for quasar light curves as a function of $\Delta T_{\mathrm{pred}}$, 
where the error bars show the 68\% confidence intervals 
evaluated from bootstrap sampling. 
The improvement in the forecasting accuracy compared to the DRW model 
grows roughly as the time separation from the last observation of the modeling part 
($\Delta T_{\mathrm{pred}}$) increases. 
Hence, at any time separation within $\Delta T_{\mathrm{pred}} \leq 500$ days, 
the AE model performs better than the DRW model in forecasting quasar light curves. 

In addition, the autoencoder trained on quasar light curves can capture
the characteristics of the DRW process. The green squares in Fig.~\ref{fig:pred_acc_deltaT} show the 
forecasting accuracy of the AE model compared to the DRW model for the simulated light curves. 
The value of $\chi_{\mathrm{red, AE}}^2 - \chi_{\mathrm{red, DRW}}^2$ for simulated light curves 
is close to zero at any $\Delta T_{\mathrm{pred}}$.  This result should be related to the fact that the autoencoder 
can recover the value of $\tau$ in the DRW process from simulated light curves as well as a fitted DRW process. 
This is impressive as it means the autoencoder succeeds in capturing the 
deterministic term in the DRW process, i.e., the exponential kernel or, at least, suggests that there is an 
autoregressive nature to quasar variability. It is the deviations in the underlying process(es) from an DRW model 
that makes the accuracy of forecasting by the AE model better than that of the DRW model.

\subsection{Visualizing the AE Features}
\begin{figure*}[t!] 
 \begin{center}
  \includegraphics[width=\linewidth]{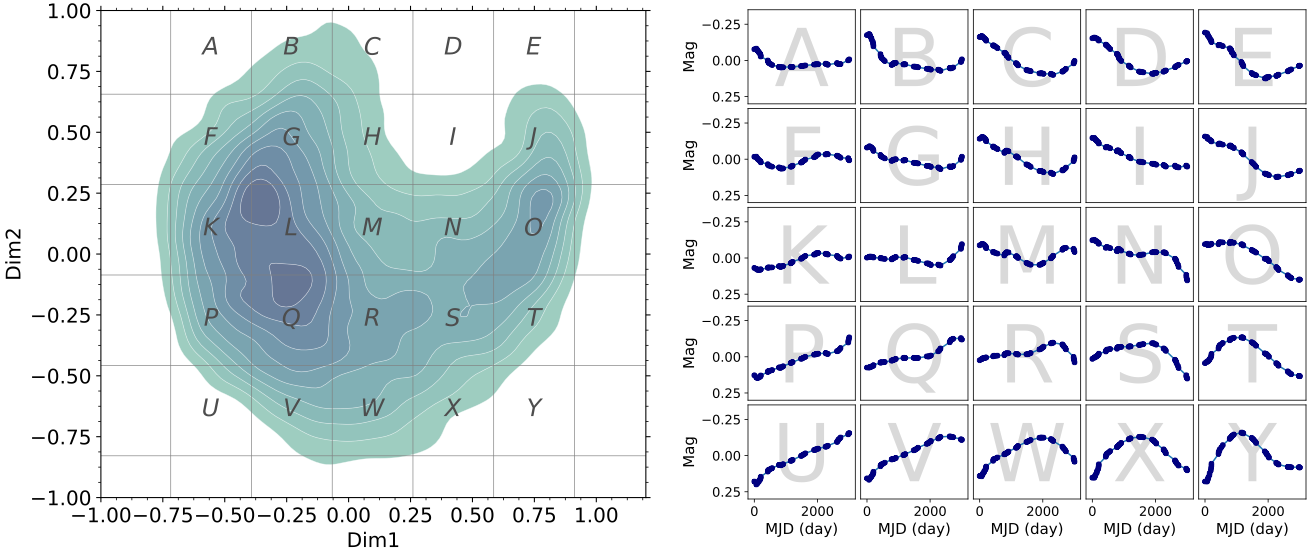}
 \end{center}
 \caption{The Gaussian kernel density estimate (KDE) of the probability density function (PDF)
 of the 16 AE features projected on two dimensional plane by PCA method (left).
 The contour levels extend from 0.9 to 0.1 in 0.1 intervals. 
 The distribution is divided into 25 pieces on the plane, A--Y, 
 and the average light curves in each piece are displayed in the right panel. 
 }  
 \label{fig:pca}
\end{figure*}
To understand what the autoencoder identifies in the quasar light curves, 
we have investigated the characteristics of the AE features. Using principal component analysis (PCA), 
the distribution of the 16 AE features can be projected onto the plane formed by the first two 
principal components (Dim1 and Dim2 respectively) as shown in the left panel in Fig.~\ref{fig:pca}.
There are three peaks in this distribution at $(Dim1, Dim2)\simeq$ $(-0.40, 0.20)$, $(-0.25, -0.10)$, and $(0.75, 0.25)$
respectively. To see what these prominent features correspond to,
we divide the distribution into a $5 \times 5$ grid labelled A--Y, and generate ``average'' light curves at each grid point using  
the decoder part of the autoencoder and an input of the averaged 16 AE features at that point. 
The resulting light curves are displayed in the right panel in Fig.~\ref{fig:pca}.  As expected, the averaged light 
curves at L, Q, and O, which roughly correspond to the peaks in the PDF, show the most global trends of temporal variability: 
namely stable, brightening, and fading, respectively. If one were to consider a polynomial expansion of the light curve, 
the three trends would be distinguishable by their primary factor, and these are the most apparent and fundamental characteristics of 
temporal variability.

On the other hand, some light curves show prominent variation over relatively short timescales, 
especially in low density regions (e.g., E, U, and Y). The shapes of such average light curves are not simple; 
they do not show a monotonic brightening/fading and their timescales/amplitudes are not symmetric. 
We may thus infer that useful information for deriving physical parameters is not associated with simple characteristics, 
such as the global trend of a light curve, as is the case with high-order coefficients in a polynomial expansion. 
However, this also shows that most quasar light curves, lying in the denser regions, do not show such prominent variation
within the observation baseline ($\lesssim$ 4,000 days). This presents a difficulty for deriving the variability timescale of quasars, 
since, qualitatively, the light curve must show at least a brightening or fading and subsequently go back to its mean value 
to estimate the time scale of the variation. 


\subsection{Information Content in Physical Parameters} \label{sec:aefeat}
\begin{figure}[!t] 
 \begin{center}
  \includegraphics[width=\linewidth]{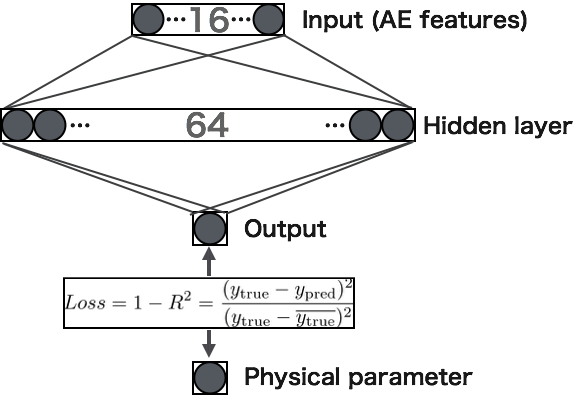}
 \end{center}
 \caption{Diagram of a multilayer perceptron (MLP) 
 for evaluating the information content ($R^2$) on physical parameters in the AE features. 
 We also apply 50\% dropout between the hidden layer and the output, 
 and normalization layer between each layer, 
 which we omit from the figure for simplicity. 
 The \texttt{relu} function is used for activation. }  
 \label{fig:mlp}
\end{figure}
The relationship between the AE features and physical parameters should be nonlinear. 
This means that the simple (partial) correlation coefficient $\rho$ or the coefficient of determination $R^2$
cannot be used directly to evaluate the information content between them. However, a multilayer perceptron (MLP)
with hidden layers can transform input in a nonlinear way, and should exploit any information in the input
associated with the physical parameter in question. We constructed a simple MLP with one hidden layer, and 
trained it to maximize the $R^2$ value between its output and a given physical parameter. 
The MLP that we used is shown in Fig.~\ref{fig:mlp}. We employed the \texttt{Adam} optimizer with a learning rate of 
$1\times10^{-4}$, and also stopped the training when the validation loss had increased with $\mathtt{patience}=250$. 
The mean value and the uncertainty (1$\sigma$) in the information content $R^2$ were 
computed with 10-fold cross-validation.\footnote{In k-fold CV, 1/k of the training set is withheld during model construction, 
and the remaining 1$-$1/k fraction of the training set is used to predict the $R^2$ of the withheld data. 
This procedure is repeated k times, with every training set source being withheld exactly once, 
so that predictions are made for each source in the training set.}
In addition, we determined the relevance of the AE features to a physical parameter 
with the following procedure: 
(1) train the MLP using all 16 AE features to maximize $R^2$ with the physical parameter, 
(2) feed the true values of one AE feature that we are interested in and zeros to the other nodes, 
(3) calculate $R^2$, and
(4) repeat this calculation (return to (2)) for each AE feature. 
This $R^2$ can be understood as the contribution of each AE feature 
to the coefficient of determination for a specific physical parameter, 
and thus can be interpreted as the relevance of it to the physical parameter under consideration. 

\subsubsection{Redshift}
In an observed light curve, the intrinsic (restframe) variation timescale is multiplied by $(1 + z)$  so we should expect a correlation 
between the observed variation timescale and redshift. \cite{Kozlowski17} has shown that previous reports of an anticorrelation 
between the variation timescale and the redshift are an artifact of insufficient temporal coverage and that any true correlation has yet to
be verified.

\begin{figure}[!t] 
 \begin{center}
  \includegraphics[width=\linewidth]{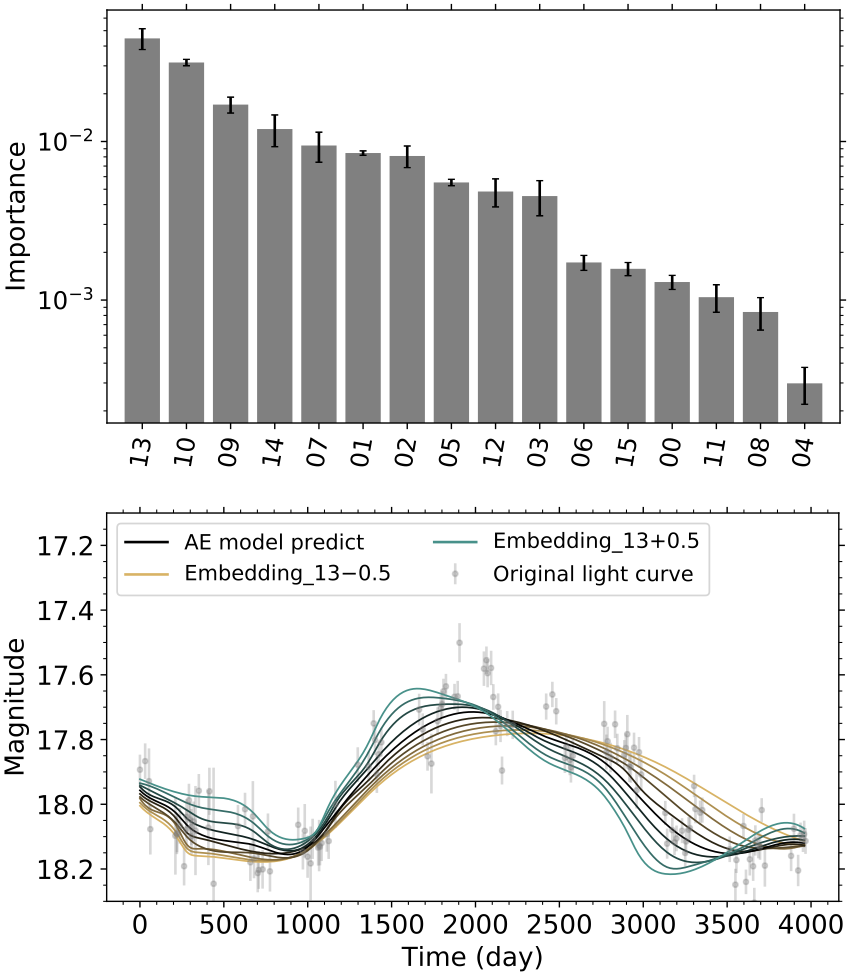}
 \end{center}
 \caption{The importance of each AE feature (top panel; see the text for detail) for the redshift.
 For illustrative purposes, the effect is shown on a sampled modeled light curve (bottom panel)
 when the most important AE feature (No. 13 for the redshift) is
 varied by $-0.5$ to $+0.5$ in $0.1$ intervals from its original value. }  
 \label{fig:red_shift_ae}
\end{figure}
The top panel in Fig.~\ref{fig:red_shift_ae} shows the importance of the 16 AE features with respect to redshift with
Feature 13 (F13) having the highest importance. To see how this feature affects the modeled light curve, 
we select a fiducial object, CRTS J110718.8$+$100417, 
whose F13 value is close to 
its mean value,\footnote{We also selected this object as there is clear brightening and fading in it and therefore the effect
of changing AE feature values is more evident.}
and vary this by $\pm 0.5$. The corresponding changes in the modeled light curve 
are shown in the bottom panel in Fig.~\ref{fig:red_shift_ae}. We see that the most significant change is
the timescale of the variation, which is precisely what we would expect, but is also the first time that such
a change has been demonstrated in quasar light curves. We note that although we have used a single object for illustrative purposes, these trends are seen in the larger statistical sample.

The variability timescale for this source (at $z = 0.633$) seems longer than the 245 day limit below which it can 
be accurately estimated for a DRW model fit. Despite a lack of a quantitative measure, though, 
the AE feature (F13) controlling the timescale of variability has a relationship with redshift.  The coefficient of determination with 
redshift is $0.07\pm 0.01$  (corresponding to the correlation coefficient $\rho = 0.3$)
implying that quasar flux variation can explain 7\% of the variance in the redshift, 
or in other words, the quasar light curve has 7\% of information content on redshift. 

\subsubsection{Optical luminosity}\label{sec:ae_lbol}
\begin{figure*}[!t]
 \begin{center}
  \includegraphics[width=180mm]{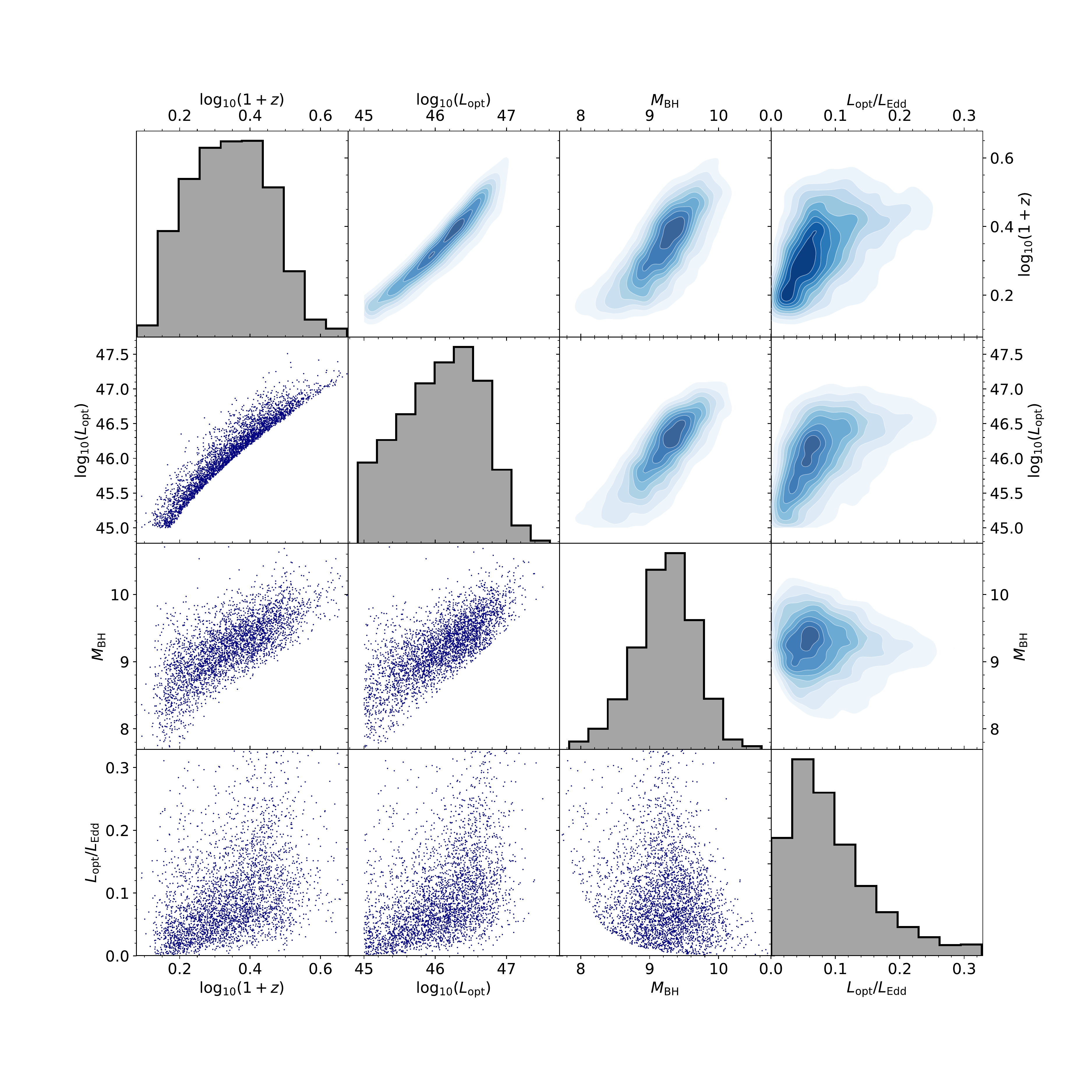}
 \end{center}
 \caption{The scatter matrix of four parameters: 
 the redshift, $L_{\mathrm{opt}}$, $M_{\mathrm{BH}}$, 
 and the ratio of the optical luminosity to the Eddington luminosity 
 ($L_{\mathrm{opt}}/L_{\mathrm{Edd}}$). 
Histograms for each parameter are shown in diagonal components. }  
 \label{fig:scat_mat}
\end{figure*}
\begin{figure}[!t]
 \begin{center}
  \includegraphics[width=\linewidth]{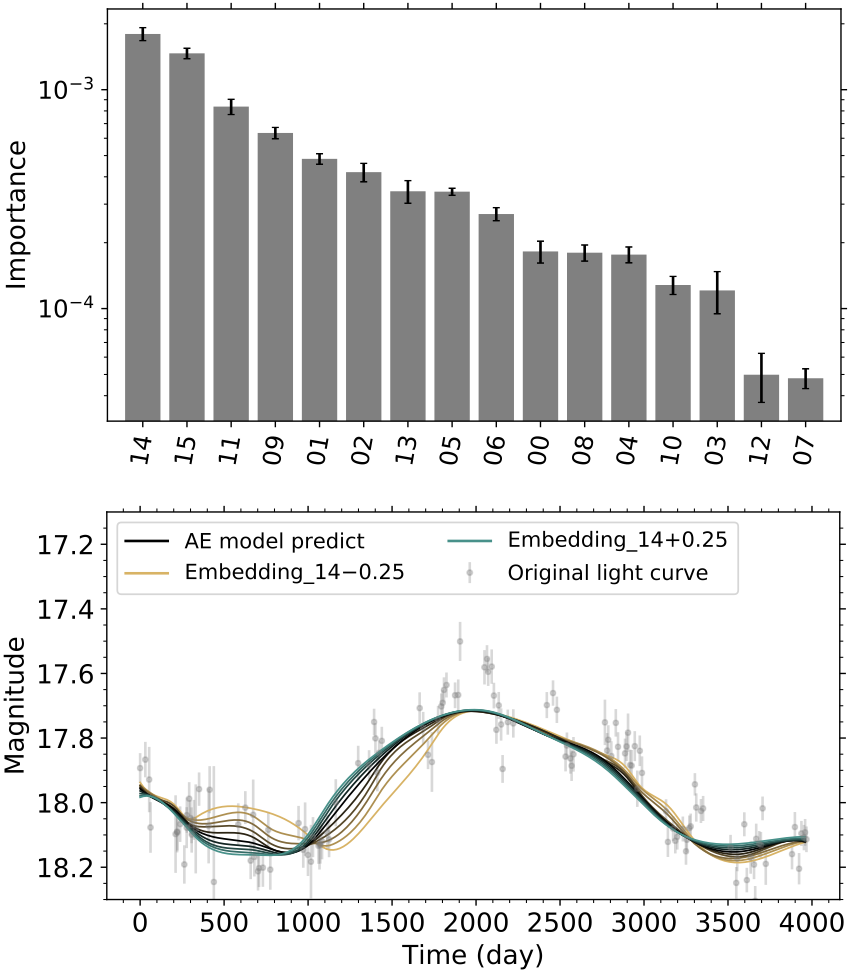}
 \end{center}
 \caption{The importance of each AE features (top panel) for $L_{\mathrm{opt}}$
 and the behaviors of modeled light curve 
 when the most important AE feature (No. 14 for the redshift) 
 varies from $-0.25$ to $+0.25$ in $0.05$ intervals from its original value. }  
 \label{fig:lbol_shift_ae}
\end{figure}

Observational biases mean that the optical luminosity, $L_{\mathrm{opt}}$, is strongly dependent on redshift
and this needs to be accounted for. Fig.~\ref{fig:scat_mat} shows the scatter matrix between six parameters
for the data set and it can be seen that the correlation between $L_{\mathrm{opt}}$ and redshift is strongly nonlinear.
We cannot, therefore, disentangle the effect of redshift on the relation between the AE features and $L_{\mathrm{opt}}$ 
with either multiple linear regression analysis or the partial correlation coefficient; instead, we include the redshift as
an additional input to the MLP alongside the AE features. AE features known to correlate with redshift, such as Feature 13 
or Feature 10, should lose their importance and other features containing information on $L_{\mathrm{opt}}$ should
emerge as more relevant. 

\begin{figure}[!t]
 \begin{center}
  \includegraphics[width=\linewidth]{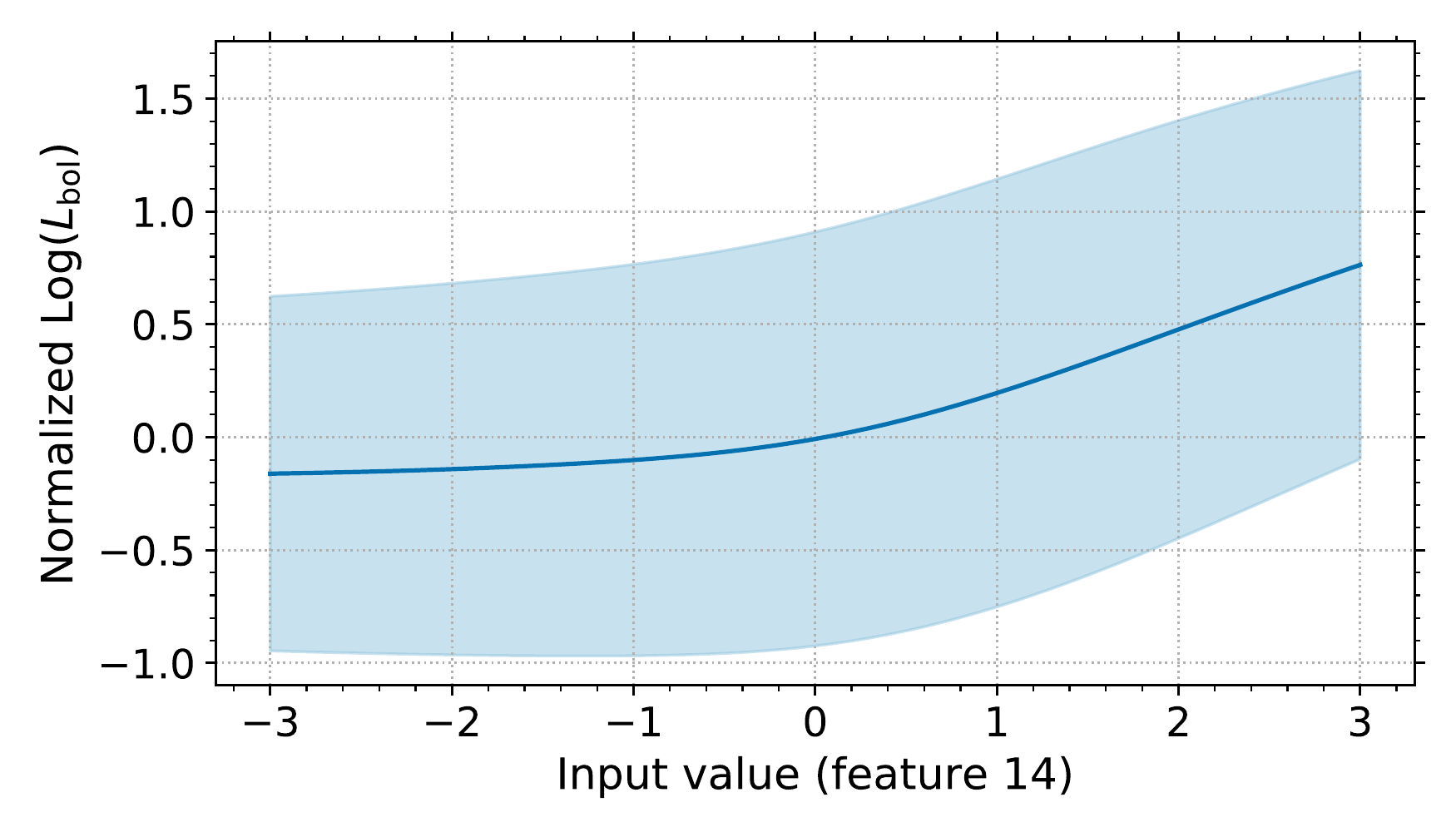}
 \end{center}
 \caption{The correlation between the input value (F14) and the normalized luminosity. 
 The average of the predicted value is shown by blue line, and the standard deviation of the prediction 
 is shown by blue shadowed region. }  
 \label{fig:lbol_feat14_corr}
\end{figure}

The top panel in Fig.~\ref{fig:lbol_shift_ae} shows the importance, $\Delta R^2$, of the AE features for 
$L_{\mathrm{opt}}$. As expected, Feature 13 and Feature 10 have lost their relevance and, instead, the most important
feature for $L_{\mathrm{opt}}$ is Feature 14 (F14). The dependence of the modeled light curve on F14 is shown in the 
bottom panel in Fig.~\ref{fig:lbol_shift_ae}, where the feature value is varied by $-0.25$ to $+0.25$ around its original value.  
The modeled light curve changes in only its brightening phase as the feature varies which suggests that the brightening 
timescale or the asymmetry of the timescale of brightening and fading relates to the luminosity of a quasar.  As shown in Fig.~\ref{fig:lbol_feat14_corr},  the output value increases as the input value to the node corresponding to F14 increases.   
It suggests a longer brightening timescale, or a higher symmetry, is possibly associated with a higher optical luminosity and
that faint quasars might tend to exhibit higher variability asymmetry and vice versa for brighter quasars.
This could indicate different physical mechanisms determining the timescale of brightening and fading. We note
that standard second-order analysis techniques, such as the power spectrum density, structure function, or correlation 
function, are not sensitive to this and neither is the DRW model. The autoencoder models the light curve itself without 
any prior assumptions and so can capture information on asymmetry if it is present.

The coefficient of determination obtained with the AE features and redshift as input is $R^2 = 0.869 \pm 0.002$ 
and with only the redshift is $R^2 = 0.864 \pm 0.002$, respectively, giving $\Delta R^2 = 0.005\pm 0.003$. 
Since the increment of the coefficient of determination $\Delta R^2$ can be understood 
as the lower limit of $R^2$ between the AE features and $L_{\mathrm{opt}}$, 
the flux variations in quasars have information on $L_{\mathrm{opt}}$. 

\subsubsection{Black hole mass}\label{sec:ae_mbh}
\begin{figure}[!t] 
 \begin{center}
  \includegraphics[width=\linewidth]{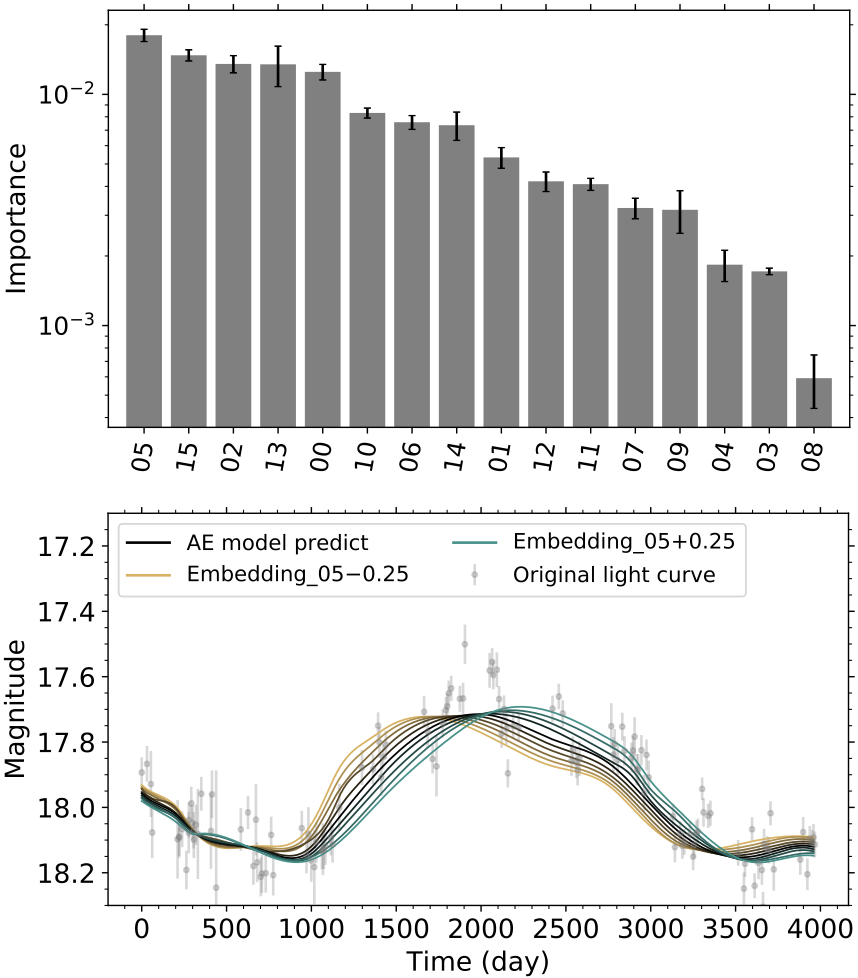}
 \end{center}
 \caption{The importance of each AE features (top panel) for $M_{\mathrm{BH}}$
 and the behaviors of modeled light curve 
 when the most important AE feature (No. 5 for the black hole mass) 
 varies from $-0.25$ to $+0.25$ in $0.05$ intervals from its original value.}  
 \label{fig:mbh_shift_ae}
\end{figure}

Fig.~\ref{fig:scat_mat} shows that the black hole mass, $M_{\mathrm{BH}}$, correlates with redshift and 
$L_{\mathrm{opt}}$, although the redshift dependency is most likely due to the strong correlation with 
$L_{\mathrm{opt}}$. We therefore include both $L_{\mathrm{opt}}$ and redshift as MLP inputs to handle these
relationships. The importance for $M_{\mathrm{BH}}$ is shown in the top panel in Fig.~\ref{fig:mbh_shift_ae} with
Feature 5 (F5) emerging as the most relevant and its effect on the modeled light curve is presented in the bottom 
panel in Fig.~\ref{fig:mbh_shift_ae}. Asymmetry in the timescale of the brightening and fading is controlled by this
 feature but in a different way to Feature 14 (see above). Since the correlation coefficient between F5 and 
 $M_{\mathrm{BH}}$ is negative ($\rho = -0.01$), the asymmetry increases as $M_{\mathrm{BH}}$ decreases, 
and this is consistent with the relation between the AE features and $L_{\mathrm{opt}}$ where the asymmetry increases 
as $L_{\mathrm{opt}}$ decreases. 

The coefficient of determination with inputs of AE features, redshift, and $L_{\mathrm{opt}}$ is $R^2 = 0.51 \pm 0.01$ 
and with only redshift and $L_{\mathrm{opt}}$, is $R^2 = 0.47 \pm 0.01$, respectively, giving $\Delta R^2 = 0.04\pm 0.01$. 
Again, the flux variations in quasars have information on $M_{\mathrm{BH}}$ because $\Delta R^2$ can be understood 
as the lower limit of $R^2$ between the AE features and $M_{\mathrm{BH}}$, as mentioned in \S\ref{sec:ae_lbol}. 
If we regard $R^2$ as the square of the correlation coefficient, we can derive the partial correlation coefficient of the AE
features with redshift, luminosity, and black hole mass. The highest partial correlation coefficient is then with 
luminosity ($\sim$ 0.1) suggesting that the AE features mainly capture characteristic variability related to luminosity
and that correlations with the other parameters might just an artifact of this relationship.

\subsection{Asymmetry in quasar light curves}
\begin{figure}[!t] 
 \begin{center}
  \includegraphics[width=\linewidth]{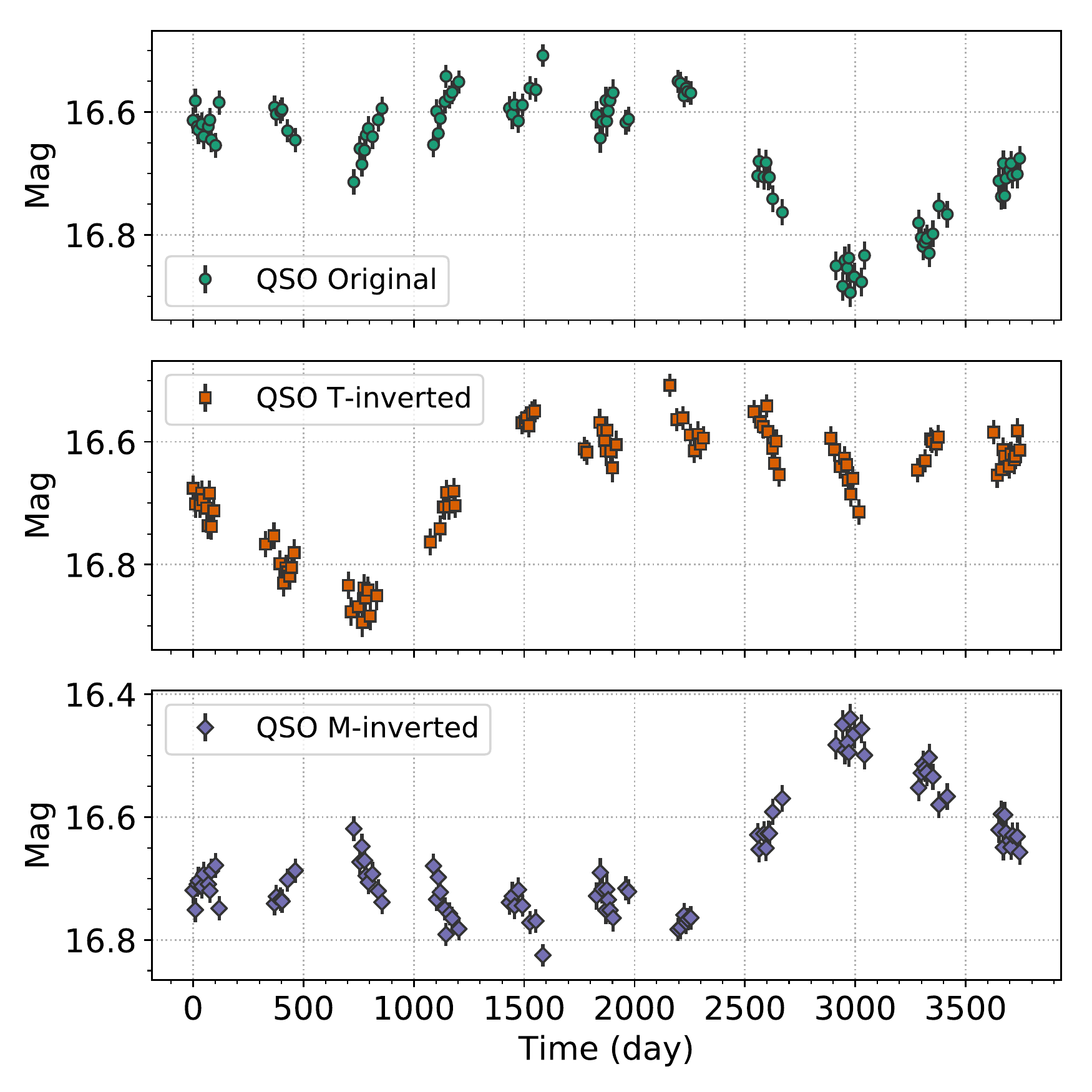}
 \end{center}
 \caption{An example of the time-inverted light curve (middle) 
 and the magnitude-inverted light curve (bottom) compared with
 the original light curve (top), respectively. }  
 \label{fig:inv_lc_example}
\end{figure}

The above results suggest that the timescales of brightening and fading in a quasar light curve are determined by 
different physical mechanisms, and that the ratio between these two components, i.e., the temporal asymmetry of 
the curve, is related to $L_{\mathrm{opt}}$. If this is the case then an autoencoder trained on quasar light curves 
would work a different way for time-inverted (T-inverted) and magnitude-inverted (M-inverted) light curves. 
An example of the T-inverted and the M-inverted light curve is shown in Fig.~\ref{fig:inv_lc_example}. 

\begin{figure}[!t] 
 \begin{center}
  \includegraphics[width=\linewidth]{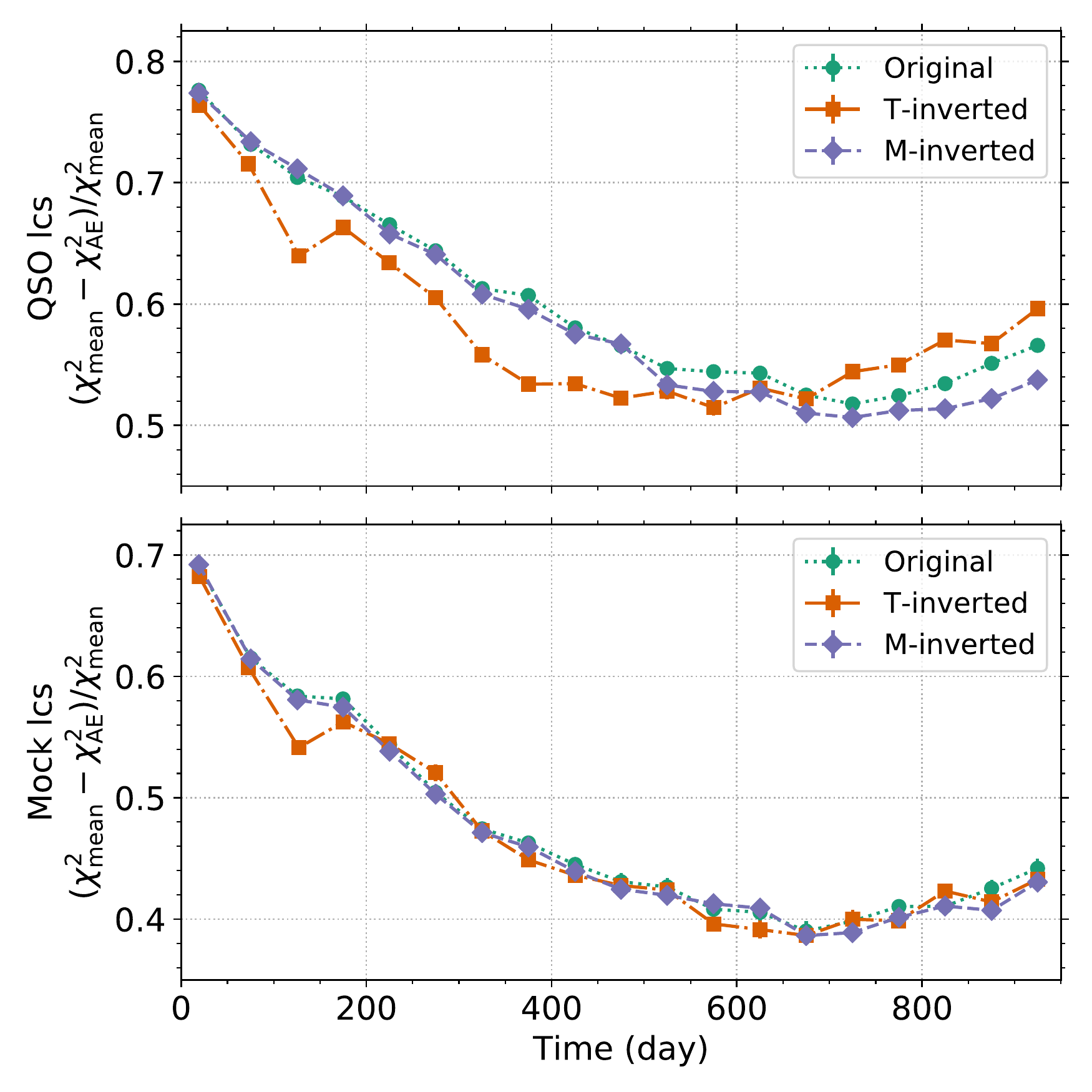}
 \end{center}
 \caption{The normalized modeling accuracy for the QSO light curves (top) and the simulated light curves (bottom). 
 Accuracies for the original light curves (green points), 
 the time-inverted light curves (denoted by T-inverted, orange squares), 
 and the magnitude-inverted light curves (denoted by M-inverted, purple diamonds) 
 are shown in each panels. }  
 \label{fig:inv_R2_m}
\end{figure}
We define the normalized modeling accuracy as: 
$
(\chi_{\mathrm{mean}}^2 - \chi_{\mathrm{AE}}^2)/\chi_{\mathrm{mean}}^2
$, where 
\begin{align}
&\chi_{\mathrm{mean}}^2 = \sum_i \left(\frac{y_i - \overline{y_i}}{e_i} \right)^2,  \notag \\ 
&\chi_{\mathrm{AE}}^2 = \sum_i \left(\frac{y_i - y_{\mathrm{pred}}}{e_i} \right)^2 . 
\end{align} 
This value is related to the coefficient of determination. Fig.~\ref{fig:inv_R2_m} shows the normalized modeling 
accuracy for quasar light curves and simulated (DRW) light curves, respectively, and in each panel, the normalized accuracy 
for the original, the T-inverted, and the M-inverted curves is displayed as a function of time in the restframe. 

The normalized modeling accuracies for the simulated light curves are almost the same among the three data, 
whereas those for the quasar light curves show different behavior. The largest deviation from the accuracy 
of the original quasar light curve comes from the T-inverted light curve; 
the accuracy for the T-inverted dataset is lower than that of the original dataset during the first half ($\sim$ 0--600 days), 
and then improves to higher than that of the original dataset after $\sim$ 700 days. 
On the other hand, the accuracy for the M-inverted data begins to slightly lag the original dataset at $\sim$ 500 days, 
and never goes to higher than that of the original dataset. The deviation from  the original dataset is smaller than for 
the T-inverted dataset.  As this behavior is not seen in the accuracies for the simulated light curves, 
it is a characteristic of the quasar light curves and not any observational bias. 

No difference in the accuracies for the three datasets of the simulated light curves is actually expected, in fact,
because the kernel in the DRW process, $\exp|\Delta t /\tau|$, is time-reversible and also brightness reversible; 
in other words, both the T-inverted DRW process and the M-inverted DRW process are still DRW processes. 
The difference in the quasar light curves thus suggests variability asymmetry is present, which is consistent with the 
results in \S\ref{sec:ae_lbol} and \S\ref{sec:ae_mbh}, indicating the existence of the arrow of time in these time series.
It is indicative that the larger discrepancy is in the accuracy for the T-inverted dataset rather than for the M-inverted 
dataset. The amplitude of the variability asymmetry is possibly small in terms of magnitude while significant in terms of time.

We note that the modeling accuracy is always larger for the quasar light curves than for the simulated light curves. 
This is probably because the autoencoder is trained only with the quasar light curves but it also
indicates that the quasar flux behavior is different from the DRW process as the parameter estimation by the 
DRW process fit is not precise. Again, the consistency in the accuracy curves for the three simulated datasets 
confirms that the discrepancy among the accuracies for the quasar datasets is not attributable to systematic effects 
such as the amount of data in each bin or seasonal observation gaps. 

\subsection{Variability Asymmetry analysis}
\begin{figure}[!t] 
 \begin{center}
  \includegraphics[width=\linewidth]{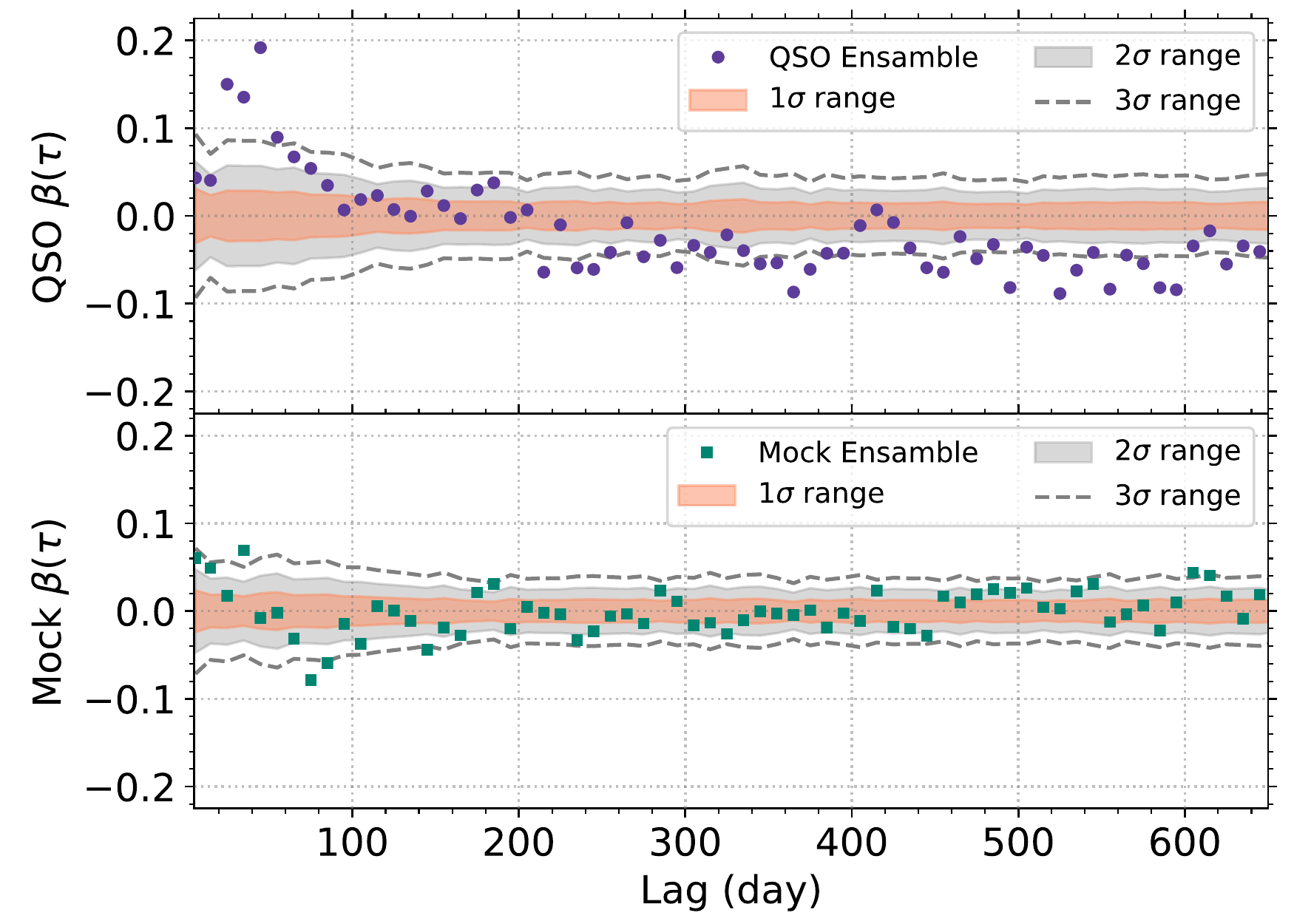}
 \end{center}
 \caption{Ensemble $\beta(\tau)$ of the QSO light curves (top) and of the simulated light curves (bottom). 
 The $1\sigma$, $2\sigma$, and $3\sigma$ uncertainty ranges of $\beta(\tau)$ are 
 shown by the orange shadowed region, gray shadowed region and dashed lines, respectively, 
 where these uncertainties are derived from the boot strapping resampling method. }  
 \label{fig:aymmetry}
\end{figure}
The results from the autoencoder model suggest the existence of variability asymmetry
in the quasar light curves. \cite{Kawaguchi98} introduced a structure function approach to estimate the 
variability asymmetry adopting two structure functions, 
$\mathrm{SF}_{\mathrm{ic}}(\tau)$ and $\mathrm{SF}_{\mathrm{dc}}(\tau)$, 
which only include pair epochs with brightening and fading flux, respectively (i.e., increasing and decreasing flux).
The asymmetry can be quantified via an asymmetry parameter $\beta(\tau)$:
\begin{align}
\beta(\tau) = \frac{\mathrm{SF}_{\mathrm{ic}}(\tau) - \mathrm{SF}_{\mathrm{dc}}(\tau)}
	{\mathrm{SF}_{\mathrm{tot}}(\tau)}, 
\end{align}
where ``tot'' refers to the total set of data pairs. 
$\beta(\tau)$ quantifies the normalized difference between the brightening and fading: 
positive $\beta(\tau)$ indicates that the light curve favors a rapid rise 
and gradual decay, and vice versa for a negative $\beta(\tau)$. Attributing quasar optical variability
to instabilities in the accretion disk (the disk instability model) produces $\beta(\tau)<0$, 
while the starburst model, which associates variability with the random 
superposition of supernovae in the starburst region of the host galaxy, yields $\beta(\tau)>0$. 
\cite{Hawkins02} also considered gravitational microlensing as a mechanism for variability and demonstrated 
$\beta(\tau)=0$ is expected in this case. 

From a sample of 401 quasars, \cite{Hawkins02} found no asymmetry signature was detected on
timescales of a year or longer. However,  \cite{Giveon99} calculated the difference between the medians 
of brightening phases and fading phases in the light curves of 42 PG quasars and found a negative asymmetry
in the variations. More recently, significant negative asymmetry was detected on a timescale longer than 300 days 
in 7,562 quasars from SDSS Stripe 82 \citep{Voevodkin11}. 

Fig.~\ref{fig:aymmetry} shows the ensemble $\beta(\tau)$ for the quasar light curves and the simulated light curves. 
To obtain the ensemble $\beta$, we calculated $\mathrm{SF}(\tau)$ for all sources in the quasar dataset, 
and then estimated using the weighted average in 10-day width bins. Since the DRW process is variability symmetric 
the ensemble $\beta(\tau)$ for the simulated light curves does not show any significant deviation 
from $\beta(\tau) = 0$. However, the ensemble $\beta(\tau)$ for the quasar light curves, presented in the top panel 
in Fig.~\ref{fig:aymmetry}, has positive $\beta(\tau)$ on short timescales, and then decreases to a statistically significant 
negative value for $\tau \gtrsim 200$ days. This behavior indicates that the brightening power of variability
is stronger than the fading power on a timescale shorter than $\sim 100$ days, while vice versa on a timescale 
longer than $\sim 200$ days.  The variability asymmetry, which is suggested by the deep learning modeling, is 
confirmed by this time domain analysis. 
 
 We note that our quasar sample has been selected to only consist of spectroscopically-confirmed sources
and that a variance-luminosity relation has been employed so that variability is solely from the accretion disk.
 Additionally, the sample size is about double that used by \cite{Voevodkin11}, and the observation cadence is 
 much denser than that of the SDSS Stripe 82 dataset. Our result is therefore the most definitive to date.
 
\section{Discussion}
\subsection{The autoencoder model and its features}

The autoencoder (AE) model we have trained on quasar light curves provides a better description of quasar optical
variability than the damped random walk (DRW) model commonly used in the literature. In particular, the forecasting
accuracy of the AE model relative to the DRW model improves as $\Delta T_{\mathrm{pred}}$ increases suggesting that
the AE model captures characteristics of the long-term behavior in quasar light curves. Quasar variability on timescales
longer than several hundred days has not been well determined so far, partly due to insufficient data sets, but also as
characteristic timescales from the DRW model are biased low for time series with temporal coverage less than ten times
the timescale in question. \cite{Caplar17} found that there are clear variations in the slopes of the quasar structure function (SF) for 
individual sources with many quasars having steeper SFs than expected from the DRW model. Quasars with higher mass 
and/or luminosity tend to have steeper power spectral density (PSD) slopes and this can be reproduced in a model where 
the PSD slope is steeper below a certain timescale which is dependent on mass and/or luminosity. This may be the same
behavior that the AE model is capturing.

The AE model is trained to reproduce all of the light curves with only 16 parameters for each object. Simple clustering analysis
of the AE features shows three populations: fading, stable, and brightening, which agrees with the most intuitive categorization. 
However, mean light curves across the grid-separated PCA projection of the features show highly flexible expressions
including global trend, variable amplitude, variable timescale, etc. This shows that the AE features have most of the latent
content of the variability but this is also tied to physical parameters since the information content on the luminosity is seen to 
increase as training proceeds and the distributions of some physical parameters on the PCA map show correlations. This
implies that the shapes of stochastic time series contain information on the physical properties and processes producing them. 

In fact, we have specifically used the AE features in deep regression models for redshift, black hole mass, and luminosity. If the
intrinsic variability is redshift independent then the observed frame light curves should show a relation between the timescale
of variability and redshift (or strictly $1 + z$). We find that AE feature 13 which controls the visual timescale of variability is the 
most relevant feature in determining redshift which validates our approach. It is perhaps more surprising, though, that an AE 
feature which controls visual asymmetry in the light curve should also be the most relevant for both luminosity and black hole 
mass. This suggests for the first time that the degree of asymmetry in quasars should be higher for low luminosity (black hole 
mass) systems which are also known to show higher amplitude variability.

\subsection{Variability asymmetry in quasar light curves}

Variability asymmetry is confirmed to be more than just a visual effect by the AE modeling and forecasting accuracies 
for time-inverted (T-inverted) and magnitude-inverted (M-inverted) light curves, i.e., the autoencoder performs differently 
for the original, the M-inverted, and the T-inverted data sets. Interestingly, accuracies for the T-inverted  
curves are higher than those for the original curves in some temporal regions, while those for the M-inverted curves are always 
lower than those of the original data set. However, as expected, these asymmetries are not seen in simulated light curves 
generated by the (time reversible) DRW process.

A more traditional time domain analysis based on the structure function also finds variability asymmetry, demonstrating that it is not 
an artifact of the deep learning approach. The SF asymmetry parameter $\beta(\tau)$ indicates that on shorter timescales 
($\tau \lesssim 100$ days) there is a shorter brightening phase with a longer fade while the reverse is seen on longer 
timescales ($\tau \gtrsim 200$ days). This trend is also consistent with the result obtained from MLP regression between 
the AE features and physical parameters where the variability asymmetry emerges in the form of a shorter brightening phase 
with a longer fading phase. 

We have seen as well that the variability asymmetry is connected to the intrinsic luminosity of quasars. Theoretical
predictions for this behavior are scant in the literature but the most plausible physical model matching our results 
is the disk instability model (\cite{Takeuchi95}; DI model hereafter) based on the concept of self-organized criticality
\citep{Bak88}. In this model, mass accretion takes place in the form of avalanches which occur only when the local mass density 
exceeds a critical value, and, simultaneously, a gradual viscous diffusion occurs regardless of the critical condition. The DI model has 
so far been mainly applied to X-ray variability in stellar mass black holes but it seems applicable to quasars with black holes
a factor of $10^{5} - 10^{8}$ larger. 
Simulated light curves generated by this process, e.g., \cite{Takeuchi95}, \cite{Kawaguchi98}, show variability asymmetry 
of $\beta(\tau) < 0$ or $\beta(\tau) >0$, depending on the avalanche rate,  
the ratio of the diffusion mass, $m^{\prime}$, to the accretion mass, $m$,
and the range of the radius of the accretion disk that we are interested in. 
\cite{Kawaguchi98} demonstrated that simulated optical light curves of quasars exhibit a negative asymmetry
on time scales of several hundred days in the rest frame which is consistent with our results. Specifically 
they found that $\beta(\tau) \sim -0.1$ is obtained with the ratio of the diffusion mass to inflow mass of 0.1--0.5. 

We can also consider the relation between variability asymmetry and luminosity within the context 
of this model. The ratio of the diffusion mass to the accretion mass controls the variability asymmetry and so
at a lower value, $m^{\prime} \sim 0.01 m$, the variability asymmetry is relatively large, $\beta(\tau) \sim 0.1$ at 
$\tau \gtrsim 100$ days, but at a higher value, $m^{\prime} \sim 0.1 m$, the asymmetry is effectively suppressed.
So luminous quasars should intrinsically have a high ratio of $m^{\prime}$ to $m$ while less luminous quasars
should have a relatively smaller value. The amplitude of variability is also suppressed by a high diffusion mass ratio
in the DI model because large amplitude variability comes from large-scale avalanches and these hardly occur when
mass diffusion is efficient. Thus a natural consequence of this is that the amplitude of variability is anticorrelated with
luminosity as has been found in several analyses.

The diffusion (or viscous) timescale for an accretion disk, $t_{\mathrm{visc}}$, gives the characteristic timescale of mass
flow and can be parameterized for a black hole of mass $M_{{\mathrm BH}}$ at $R \sim 150r_g$ \citep{Stern18} as:

\begin{align}
t_{\mathrm{visc}} \sim 400 \, \mathrm{yr} \left( \frac{h/R}{0.05}\right) ^{-2} \left( \frac{\alpha}{0.03} \right)^{-1} \left(
\frac{M_{{\mathrm BH}}}{10^8 M_{\sun}} \right) \left( \frac{R}{150r_g} \right)^{3/2}
\end{align}

\noindent
where $\alpha$ is the disk viscosity parameter, $h/R$ is the disk aspect ratio, $R$ is the disk radius, and 
$r_g = GM_{{\mathrm BH}}/c^2$ is the gravitational radius. $t_{\mathrm{visc}}$ should be inversely proportional
to the amount of diffusion mass, $m^{\prime}$, per unit time and so  $\dd m^{\prime}/\dd t \propto \alpha(h/R)^2$,
which should be higher for luminous quasars. As both the amount of diffusion mass and the inflow mass per unit
time should increase simultaneously as $\alpha$ increases, the ratio $m^{\prime}/m$ should be fairly independent
of $\alpha$. The scale height, $h/R$, would thus be the most plausible physical parameter responsible for 
differences in the variability asymmetry in the standard disk regime.
 
One possible explanation is that quasars with higher metallicity (based on the measured metallicity of the broad line 
regions) appear to have systematically smaller continuum reverberation lags, i.e., smaller disk sizes. \cite{Jiang17} 
found that high-luminosity quasars seem to follow a disk temperature profile, $T(R) \propto R^{-1/\beta}$, 
with $\beta < 4/3$, which is also confirmed by microlensing \citep{Blackburne11, Hall14}. If high-luminosity quasars have 
a high volume of metallicity resulting in a small emission region for the optical band, then a larger scale height, $h/R$,
can be expected for a fixed height disk at radius $R$. In fact, a relation between black hole mass and quasar metallicity
has already been suggested, e.g., \cite{Warner03, Kisaka08}. The relationship between disk size and metallicity may 
result in large changes in disk opacity as a function of the gas metallicity, which can significantly alter the thermal
properties and structure of the accretion disk. This might then explain the connection between variability asymmetry and
luminosity. Alternatively, a large mass accretion rate can also be responsible for a large scale height as it should lead to 
a large amount of photon emission from the disk and also a large surface density. The gas pressure at the radius
exhibiting a fixed temperature should thus be relatively larger for luminous quasars, and, as the gas pressure contributes
to the scale height. luminous quasars should have accretion disks with a relatively larger scale height.

However, our variability asymmetry is positive, $\beta(\tau) > 0$, on a short timescale ($\tau < 100$ days) and the 
opposite on longer timescales ($\tau > 200$ days). A single physical mechanism may be responsible for variability 
asymmetry on all timescales or different mechanisms may produce it on the short and longer timescales respectively.
\cite{Takeuchi95} showed that simulated and observed X-ray fluctuations at a radius of $\sim 3000 r_g$ exhibit 
positive variability asymmetry, which supports a single mechanism, but the simulations of \cite{Kawaguchi98} consider
a different radius range. As the disk temperature decreases proportional to $\sim R^{-3/4}$, the radius range
emitting higher energy photons should be smaller relative to that producing lower energy photons which might 
mean that short timescale fluctuations from the smaller region show the positive variability asymmetry and vice versa
for the longer timescale fluctuations. The starburst model, which attributes aperiodic luminosity variations to the
random superposition of supernovae in the nuclear region, would be consistent with this and \cite{Kawaguchi98} 
demonstrated that it produces significant positive asymmetry on a timescale of 1--100 days in agreement with our 
results. Additionally, a high supernova rate implies larger luminosity quasars and lower variability amplitude.
This model cannot, however, explain the negative asymmetry seen and so another process must be responsible
for the transition seen from positive to negative asymmetry as the variation timescale increases.

\begin{figure*}
 \begin{center}
  \includegraphics[width=\linewidth]{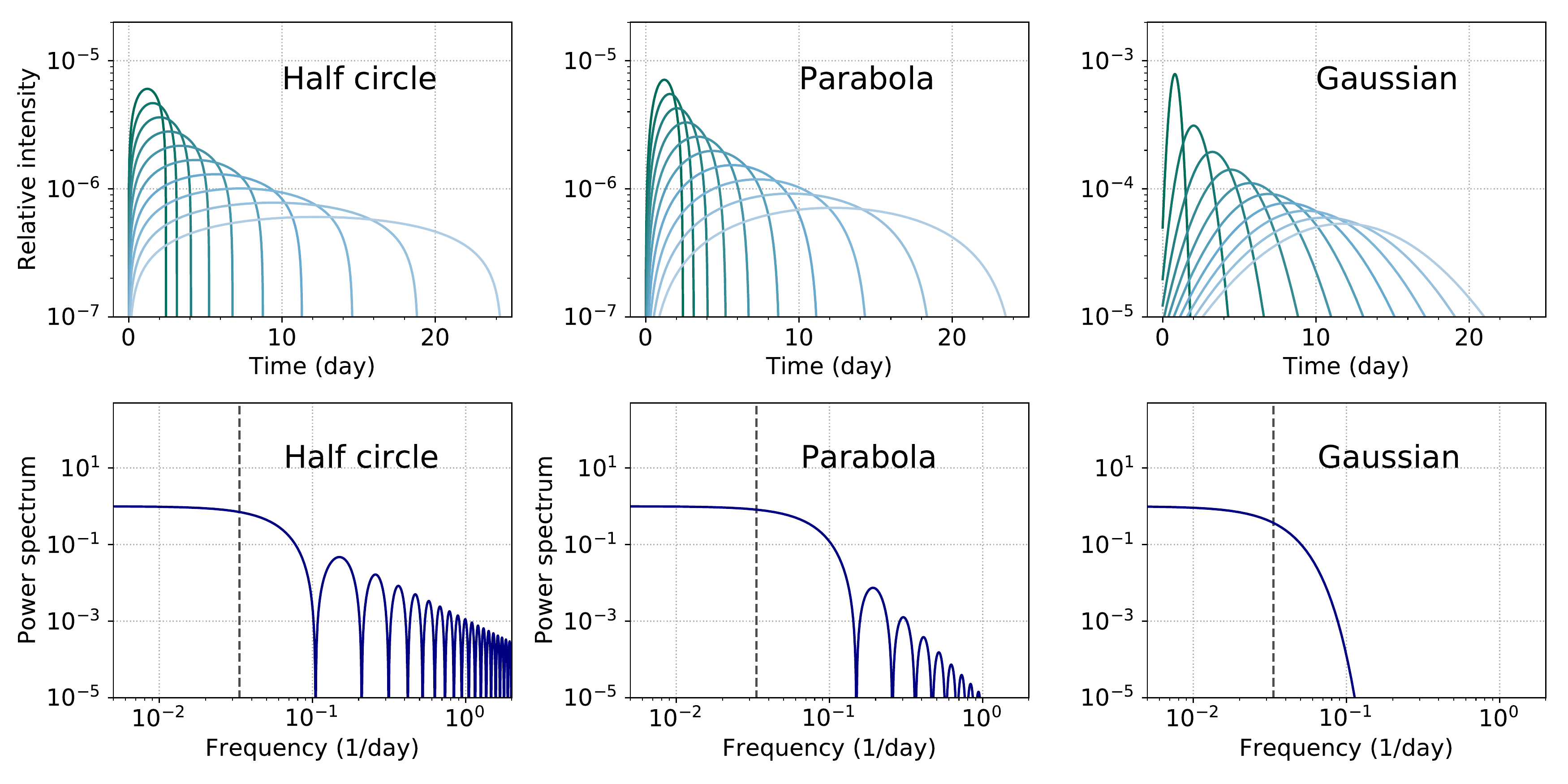}
 \end{center}
 \caption{Examples of simple kernel functions (top three panels) and their PSDs (bottom three panels). Ten typical timescales
 are shown for each kernel function and the dashed line in the bottom panels shows the typical frequency corresponding
 to the typical timescale. }  
 \label{fig:psd_kernel}
\end{figure*}

It is possible that the variability seen is not the direct product of a single intrinsic process but a convolution of several.
The optical flux of quasars must contain broad line emissions which are thought to be produced $\sim 10-100$ light days
from the central region (e.g., \cite{Peterson97}). Although the contribution to the total flux is only of order a few percent, it is 
detectable in statistical measures of variability, such as the autocorrelation function. 1 -- 50 day continuum reverberation lags
in the UV-optical bands have also been measured in several local AGN, e.g., NGC 4395, NGC 4593, NGC 5548, and NGC 
4151 (see \cite{McHardy18} and references therein). Light curves with contributions from both phenomena can be 
produced by convolving the underlying process with an appropriate kernel and Fig.~\ref{fig:psd_kernel} shows three example
kernel functions and their PSDs. In principle, any kernel will reduce high-frequency power, resulting in a steeper PSD spectral
index than the original\footnote{The PSD of convolved time-series can be calculated by $P(f)\times\Phi(f)$, 
where $P(f)$ is the original PSD and $\Phi(f)$ is the PSD of the kernel function. }. 

\begin{figure}
 \begin{center}
  \includegraphics[width=\linewidth]{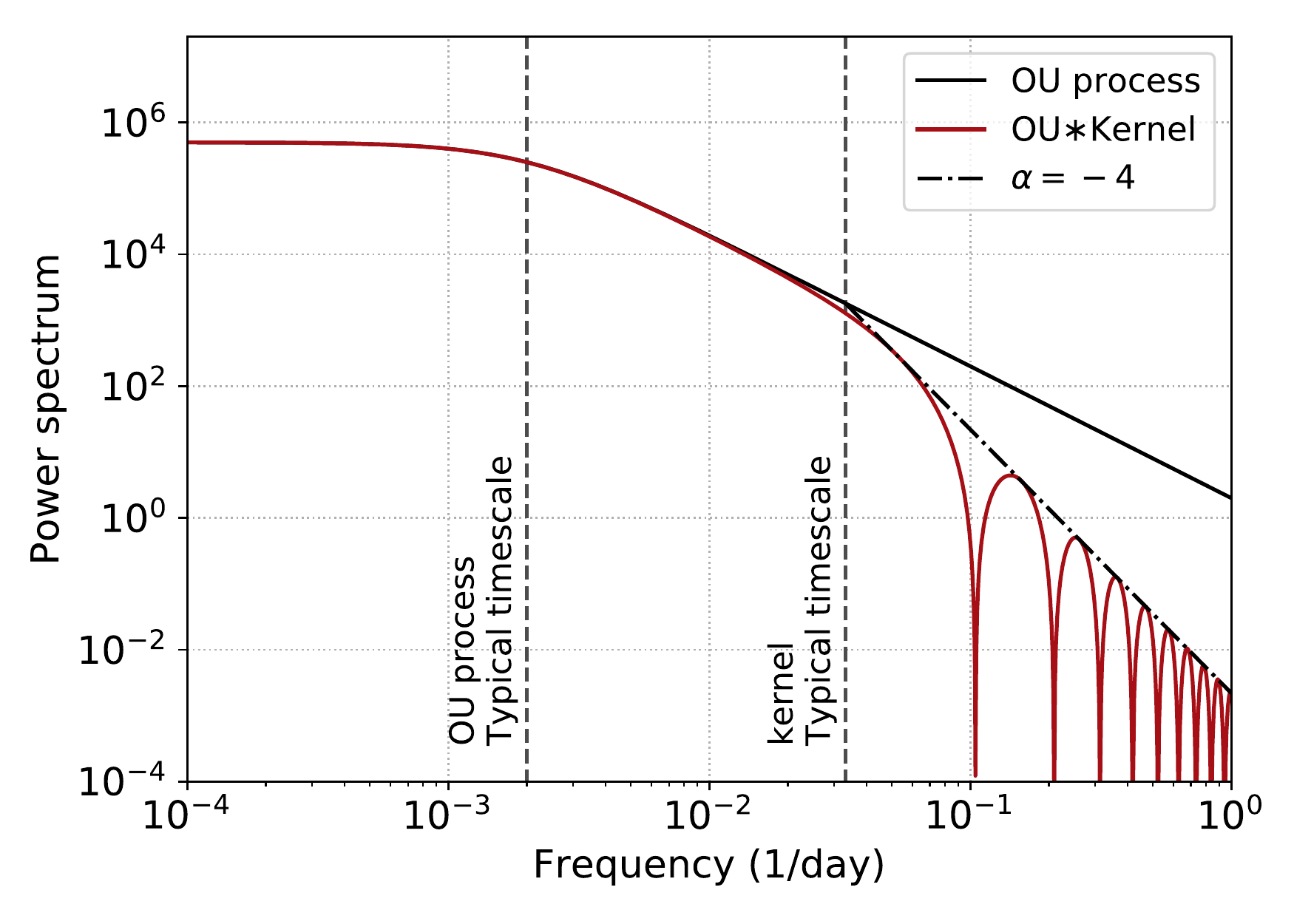}
 \end{center}
 \caption{The PSD of the DRW process convolved by the half circle kernel. 
 The original (the DRW process) PSD is shown by the black solid line, 
 while the convolved PSD is shown by the red solid line. 
 The power-law function with the index $\alpha = -4$ is also represented by dot dashed-line 
 as a reference. }  
 \label{fig:psd_OU_kernel}
\end{figure}
Fig.~\ref{fig:psd_OU_kernel} shows an example of the PSD of the DRW process convolved with a kernel.  At frequencies
above the typical timescale of the kernel function, the PSD shows a steeper spectral index ($\alpha=-4$ with this kernel) 
than that of the original PSD ($\alpha=-2$ for the DRW process). We can thus expect at least two breaks in the PSD when 
convolved with a kernel function with steepened spectral indices as the frequency increases. The light curve from the convolved process
should also show a higher correlation coefficient than that of the original temporal flux variation and the correlation should have a duration
roughly corresponding to the typical timescale of the kernel. The observed quasar PSD slope is significantly steeper than $\alpha = -2$ 
on timescales shorter than $\sim 1$ month [REF]. Kernel convolution naturally generates the steeper PSD slope above the 
typical timescale of the kernel, which should itself correlate with black hole mass and/or quasar luminosity, resulting from the scaling law 
with $M_{\mathrm{BH}}$. This suggests that assuming that quasar flux variation contains some amount of reverberated flux can 
explain the complex behavior of the time variation and also the PSD characteristics revealed so far. The existence of a kernel function, 
which manifests as the timescale with a high correlation coefficient in the quasar variation, also possibly explains the higher 
modeling/forecasting accuracy of the autoencoder model for the T-inverted data set than that for the original data set. 

\section{Summary and Conclusions}

We have constructed a nonparametric model to describe the optical variability of quasars with a small number of representative
features using a recurrent autoencoder (AE), a type of deep neural network suited for time series (sequential) data. The AE has been 
trained to both model (predict) and forecast quasar behavior by using truncated time series (by 500 days) as 
input and minimizing the reduced chi squared between the output of the network and the original full light curve.
With real data, it provides a more accurate forecast than the corresponding damped random walk (DRW) model fit to the input and the 
AE performance improves relative to the DRW model with increasing forecasting time. With simulated light curves from a DRW process, however, both models show comparable accuracy and this demonstrates that the trained AE can capture properties in the
DRW process and, indeed, recover DRW process parameters. It also shows that quasar variability differs from a DRW model. 

The AE also provides a compact learned representation of the input data set (and thus quasar variability) via the encoded features from the most compressed hidden layer. These enable investigations of the relations between the temporal flux variation of quasars and their physical
parameters, specifically redshift, luminosity, and black hole mass. To simplify this, we trained a multilayer perceptron (MLP) model on the 
AE features to maximize the coefficient of determination ($R^2$) between the respective physical parameter and the output of the MLP.
The importance of each AE feature was also evaluated based on its effect in improving $R^2$. The feature responsible for the timescale
of variability was found to be the most relevant for redshift, as expected; however, we also identified the feature controlling variability asymmetry
as the most important for predicting luminosity and black hole mass. 

The existence of variability asymmetry is shown by different model/forecasting accuracies for time-inverted (T-inverted) and magnitude-inverted (M-inverted) versions of the input data set. This is not seen when dealing with simulated time series from a DRW process which is naturally time symmetric. The AE fit to the T-inverted data set shows a higher forecasting accuracy than the original dataset over a limited time range and this implies that the T-inverted light curves have information on future variability which equates to past variability in the original data set. Independent analysis of the same data sets using the structure function confirms a variability asymmetry connected with optical luminosity and black hole
mass and that the hysteresis in the variability differs from a DRW process. A positive variability asymmetry is present on short timescales ($\lesssim 100$ days) and a negative asymmetry on longer timescales ($\gtrsim 200$ days). 

The observed asymmetry is consistent with the disk instability (DI) model where variability from the accretion disk is ascribed to instabilities in the disk as matter flows and the asymmetry to large-scale avalanches. Light curves generated from Monte Carlo simulations of this behavior show a positive asymmetry in variability from a compact inner region close to the central black hole while a negative asymmetry in the variability emerges
 from a wide outer region \citep{Takeuchi95, Kawaguchi98}. CRTS is an unfiltered survey and therefore sensitive from $\sim$ NIR to UV wavelengths so the 
 observed temporal flux variations should contain those originating over a wide range of the accretion disk. Given the disk temperature 
 profile as a function of radius, $T \propto R^{-3/4}$, fast variability should come from the inner compact region, possibly generating the positive
 asymmetry, and vice versa for longer timescale variability.

The magnitude of the variability asymmetry is controlled by the ratio of the diffusion mass to the inflow mass 
with the asymmetry diminishing as the ratio increases. We found that the asymmetry decreases as the luminosity increases which requires efficient mass diffusion in the accretion disks of luminous quasars. This can be interpreted in light of prior observational results that quasars with higher metallicity have smaller disk sizes at a fixed wavelength and also that black hole mass seems to correlate with metallicity. If we can assume that the height of the disk is determined by the mass accretion rate at a radius and that the dispersion of the mass accretion rate is small among quasars, then luminous quasars should have a relatively smaller disk and larger scale height ($h/R$) at a portion of the accretion disk with fixed disk temperature. Alternatively, the high accretion mass rate could be responsible for a higher accretion disk scale height in luminous quasars. As the scale height is proportional to the diffusion mass rate, luminous quasars should have lower variability asymmetry with lower variability amplitude as the efficient mass diffusion results in fewer large avalanches. This is consistent with previous results in the literature that luminous quasars exhibit lower amplitude variability.

The AE fit to the T-inverted data set shows a lower modeling/forecasting accuracy over particular timescales. A natural interpretation is that quasar variability retains information from prior activity over certain time frames. This can be represented mathematically by treating an observed time
series as the convolution of an intrinsic time series and a kernel function. Contributions from reverberation at the broad line region and/or from the accretion disk itself must be present in the quasar flux variation and these typically show $\sim$10 -- 100 day time lags relative to the intrinsic flux variability and so are a predictable component with such timescales. The kernel function can also address the discrepancy from the DRW process in the quasar PSD: the slope of the PSD gets steeper on shorter timescales and the slope seems to correlate with luminosity and/or black hole mass. The kernel convolution significantly reduces variability on a timescale shorter than that of the kernel function, and the typical timescale should scale
with black hole mass. 

Finally, the next generation time domain surveys, such as the Zwicky Transient Facility (ZTF; \cite{Bellm2019, Graham2019}) and the Vera Rubin Observatory \citep{Ivezic2008}, will provide multicolor observations with only a few night's cadence over several years for millions of quasars. This will greatly improve our ability to test and assess explanations of quasar physics; for example, our model for quasar variability suggests different variation characteristics in different energy bands. The disk instability model predicts that bluer color on a short timescale variability changes to redder color on longer timescales and the negative variability asymmetry on longer timescales should be smaller in higher energy bands. It is also interesting to consider what an autoencoder trained on multicolor higher cadence photometric observations might show. Physical labels from spectra, such as the equivalent width of an emission line or the strength ratio of certain emission lines, can be mapped to the projected distribution of autoencoder features provided by PCA or other dimensional reduction techniques, such as T-SNE or UMAP. This would provide a novel picture clearly relating spectroscopic properties to variability characteristics. 

\if(0)
\subsubsection*{Data selection}
QSO light curves were originally taken from the QSO archive of Catalina Real-Time Transient Survey (CRTS) project. 
In order to investigate \textit{pure} temporal flux variations originated in the accretion disk, 
we selected sources with high-variation and high-luminosity 
so that we can avoid the affect from host galaxy contamination, 
which potentially distorts the results from a systematic analysis.
The resulting sample is $\sim$ 15,000 QSO light curves 
with an observational baseline ranging $\sim$7--11 yrs with $\sim$10--30 days cadence. 

On the other hand, relatively high autocorrelation coefficients on the typical variability timescale of $\sim60$ d
is revealed through the ensemble correlation analysis. 
It may be related to the hysteresis suggested by the analysis with the T-inverted dataset. 

For a short timescale variability, 
QSO light curves obtained by ongoing space-based optical timing missions like 
\textit{Kepler}, \textit{K2}\cite{Howell14}, and \textit{TESS}\cite{Ricker16}  
will uncover that QSO varies at an astounding diversity of timescales and amplitudes, 
possibly attributed to the size or the geometry of the accretion disk or the broad line emission region,  
that with the confluence of ground-based surveys in the optical band.
The method to estimate the kernel function will generate fruitful results. 
\fi

\section*{Acknowledgements}

This work was supported in part by the NSF grants AST-1518308, and AST-1815034, and the NASA grant 16-ADAP16-0232. The work of DS was carried out at Jet Propulsion Laboratory, California Institute of Technology, under a contract with NASA. 

YT was funded by JSPS KAKENHI Grant Numbers JP16J05742. YT studied as a Global Relay of Observatories Watching Transients Happen (GROWTH) intern at Caltech during the summer and fall of 2017. GROWTH is funded by the National Science Foundation under Partnerships for International Research and Education Grant No 1545949.

NK acknowledges the support by MEXT Kakenhi Grant number 17H06362
and the JPSP PIRE program.

\bibliographystyle{aasjournal}
\bibliography{autoencoder}



\end{CJK*}
\end{document}